\documentclass{aa}

\usepackage{graphicx}
\usepackage{txfonts}
\usepackage[dvipsnames]{xcolor}
\usepackage{hyperref}
\hypersetup{
    colorlinks=true,
    linkcolor=blue,
    citecolor=blue,
    filecolor=blue,
    urlcolor=blue
}
\usepackage{subcaption}
\newcommand{\rveil}{\texorpdfstring{$r_{\text{veil}}$}{r_veil}}
\newcommand{\domi}{\citetalias{domi}}
\newcommand{\cinn}{\citetalias{cinn2}}
\newcommand{\guarc}{\citetalias{guarcello}}
\newcommand{\gaia}{\citetalias{gaia2}}
\newcommand{\teff}{\texorpdfstring{$T_\mathrm{eff}$}{t_eff}}
\newcommand{\Av}{\texorpdfstring{$A_\mathrm{V}$}{A_V}}
\newcommand{\logg}{\texorpdfstring{$\log {g}$}{logg}}
\newcommand{\go}{\texorpdfstring{$G_\mathrm{0}$}{G_0}}

\begin{document}

   \title{External photoevaporation in the center of Tr14}

   \author{K. Gkimisi,
          \inst{1, 2}
          \and
          D.E. Kang \inst{1}
          \and
          D. Itrich \inst{3, 4, 5}
          \and 
          R. Anania \inst{6}
          \and 
          E. Fiorellino \inst{1,7}
          \and
          S. Gavino \inst{1}
          \and
          R. S. Klessen \inst{8,9}
          \and
          V. F. Ksoll \inst{8}
          \and
          G. Milazzo \inst{10, 11}
          \and
          A. F. McLeod \inst{12, 13}
          \and
          T. Preibisch \inst{4}
          \and
          V. Roccatagliata \inst{1,14}
          \and
          L. Testi \inst{1,14}
          \and
          C. Toci \inst{15}
          }

   \institute{
            Alma Mater Studiorum Università di Bologna, Dipartimento di Fisica e Astronomia (DIFA), Via Gobetti 93/2, I-40129, \\Bologna, Italy \\
            \email{katia.gkimisi@unibo.it} 
        \and
            INAF – Osservatorio di Astrofisica e Scienza dello Spazio, Via P. Gobetti 93/3, 40129 Bologna, Italy
        \and
            European Southern Observatory, Karl-Schwarzschild-Str. 2, 85748 Garching bei M\"{u}nchen, Germany
        \and
            Universit\"{a}ts-Sternwarte, Ludwig-Maximilians-Universit\"{a}t, Scheinerstrasse 1, 81679 M\"{u}nchen, Germany
        \and
            Steward Observatory, The University of Arizona, Tucson, AZ 85721, USA
        \and
            School of Physics, Trinity College Dublin, Dublin, Leinster, IE
        \and
            INAF - Osservatorio Astronomico di Trieste, via Tiepolo 11, I-34143 Trieste, Italy
        \and
            Universit\"{a}t Heidelberg, Zentrum f\"{u}r Astronomie, Institut f\"{u}r Theoretische Astrophysik, Albert-Ueberle-Stra{\ss}e 2,D-69120 \\Heidelberg, Germany
        \and
            Universit\"{a}t Heidelberg, Interdisziplin\"{a}res Zentrum f\"{u}r Wissenschaftliches Rechnen, Im Neuenheimer Feld 205,D-69120 \\Heidelberg, Germany
        \and
            Dipartamento di Fisica e Chimica - Emilio Segrè, Università degli Studi di Palermo, Palermo, Italy
        \and
            INAF - Osservatorio Astronomico di Palermo “Giuseppe Salvatore Vaiana”, Palermo, Italy
        \and
            Department of Physics, Centre for Extragalactic Astronomy, Durham University, Durham, UK
        \and
            Department of Physics, Institute for Computational Cosmology, Durham University, Durham, UK
        \and 
            INAF-Osservatorio Astrofisico di Arcetri, Largo E. Fermi 5, I-50125, Firenze, Italy
        \and
            Departamento de Fisica aplicada III, ETSI Universidad de Sevilla, Camino de los Descubrimientos, 41092 Sevilla
            }
   \date{Received August 7, 2026; accepted September 15, 2026}

 \abstract{Most stars form in massive stellar clusters, where intense far-ultraviolet (FUV) radiation from OB stars can accelerate the dispersal of protoplanetary disks through external photoevaporation. Quantifying the impact of this process on disk evolution remains challenging, particularly in distant and crowded star-forming regions.}{We investigate the low-mass stellar population in the highly irradiated center of Trumpler 14 and examine how the local FUV radiation field influences circumstellar disk evolution.}{We extracted stellar spectra from VLT/MUSE observations using an enhanced background-optimized extraction method, carefully corrected for instrumental response as a function of wavelength and finally derived the stellar parameters with the SAPSAL deep-learning framework. We combined these results with a probabilistic estimate of the local FUV field and near-infrared photometry to identify disk-bearing stars.}{We identify 310 bona fide cluster members with an age of $0.66^{+0.73}_{-0.35}$ Myr and local radiation fields spanning $\log(FUV)\simeq4.2-6.0\ G_{\mathrm{0}}$.}{The fraction of stars exhibiting near-infrared excess decreases from 30\% at the lowest FUV fluxes to under 5\% at $\log(FUV)\gtrsim5.6 \ G_{\mathrm{0}}$, while no significant correlation is found between our simplified estimate of the optical veiling and the local FUV field. The pronounced decline in the disk fraction with ambient FUV flux provides strong evidence for external photoevaporation rapidly dispersing circumstellar disks in the core of Trumpler 14. Our results confirm previous disk dispersal estimates and extend to higher values of the FUV field strength.}

   \maketitle
%
\defcitealias{gaia2}{GaiaDR3}
\defcitealias{cinn2}{Kang25}
\defcitealias{domi}{Itrich24}
\defcitealias{guarcello}{Guarc23}

\section{Introduction \label{sec:intro}}

Planets form within protoplanetary disks of gas and dust that surround young stellar objects (YSOs) in star-forming regions (SFRs). The evolution of these disks determines the reservoir of material available for planet formation and therefore directly sets the mass budget, orbital architecture, and atmospheric composition of the resulting planetary systems \citep[e.g.,][]{williams2011, parker}. In stellar clusters, far-ultraviolet (FUV) and extreme-ultraviolet (EUV) radiation from OB-type members can heat and disperse the outer regions of neighboring protoplanetary disks through a process known as external photoevaporation \citep[e.g.,][]{johnstone1998, scally2001,wh}. For low-mass stars, the shallow gravitational potential in the outer disk cannot retain the FUV heated gas, driving disk dispersal \citep{haworth2018a}. By altering the disk structure and the distribution of material, this process can profoundly affect the final stellar mass, types and locations of the resulting forming planets \citep{winter2022, qiao2023,qiao2026,huang2024a} as well as planet formation efficiency \citep{ndugu}. Given the potential role of external photoevaporation in shaping star and planet formation, and the large fraction of stars that form in clustered environments, it is essential to observationally constrain its impact on protoplanetary disk evolution in such regions.

The first observational evidence for external photoevaporation came from Hubble Space Telescope images of the Orion Nebula cluster \citep[ONC;][]{proporion1,odell1994, mccaugh1996, proporion2}, where more than 150 objects, known as \textit{proplyds}, were revealed. Proplyds are defined as circumstellar disks surrounded by a comet-like shape with the tail oriented away from the UV source, a morphology produced by an externally induced photoevaporative wind \citep{wh}. Since their discovery, irradiated objects have been identified in several other regions (e.g., proplyds in NGC 1977 and NGC 2024 \citealp{Kim2016,Haworth2021}, respectively, and evaporating globules in Tr14 \citealp{mesa2016}), though the vast majority of well-characterized proplyds remain in the ONC. 

However, it is now well established that most stars, and thus most planetary systems, form in environments that are denser, more massive, and more energetically hostile \citep[e.g.,][]{miler,ladalada,krumholz, wh}, typically located at significantly greater distances. For example, while the ONC \citep[$\sim 400 \text{ pc}$,][]{megeath2012, wright2020, aru} contains only one primary UV source irradiating the region \citep{odell2017}, many regions at larger distances, such as Cygnus OB2 \citep[$\sim$ 1.4 kpc,][]{guarc2013,guarcello}, Trumpler 14 \citep[$\sim$ 2.35 kpc,][]{preibis1, preibis2, preibis3, shull2021,goppl2022}, Westerlund 1 \citep[$\sim$ 4.2 kpc,][]{clark2005, negeruela2022}, and NGC 3603 \citep[$\sim $7 kpc,][]{harayama2008}, host an abundance of OB-type stars. These rich regions offer an environment where the effect of external photoevaporation is expected to be relevant for a larger volume within the cluster and thus a larger number of stars.

In this work we study Trumpler 14 (Tr14), one of the youngest \citep[$\leq 1$\,Myr;][]{penny1993, vasquez1996, degioia2001, carraro2004, domi} and densest open clusters in the Carina Nebula Complex (CNC). Tr14 shows low interstellar extinction \citep[$\sim 2.10\ \text{mag}$,][]{walborn1995, hur2023} and hosts $\sim20$ O-type stars \citep{shull2021, berlanas2023} with 5 of them located at the core of the cluster. This extreme environment makes the core of Tr14 an ideal candidate for studying the environmental effect of massive stars on the evolution of their low-mass neighbors. 

While solar-like to massive stars in Tr14 have been extensively studied \citep[e.g.,][]{degioia2001,wfi, damiani2017}, only two recent works, \citet{domi} and \citet{cinn2} (hereafter \domi{} and \cinn{}, respectively) have explored the possibility of using integral field intermediate resolution optical spectroscopy to develop a methodology to fully characterize the lower-mass ($<1\mathrm{M_{\odot}}$) stellar population and the disk-star interaction. \domi{} presented a spectroscopic census of 780 low-mass sources based on observations of the entire cluster obtained with the Multi Unit Spectroscopic Explorer (MUSE) integral-field unit \citep[IFU,][]{muse} on the Very Large Telescope (VLT), excluding stars with uncertain photometry. Stellar properties were derived through comparison with spectral templates of Class III stars from \citet{manara2013, manara2017}. The cluster age was estimated as $\sim 1\text{ Myr}$, with the stellar masses ranging from $0.17\,\mathrm{ M_{\odot}}\ \text{to }3\,\mathrm{ M_{\odot}}$. Their work provided the first comprehensive spectroscopic characterization of the low-mass population in the CNC and established the basis for further studies of the stellar population in Tr14.

In addition, \citet{cinn}, \cinn{} developed the deep learning framework SAPSAL\footnote{\href{https://github.com/kangdaeun/SAPSAL}{https://github.com/kangdaeun/SAPSAL}} for inferring the effective temperature (\teff{}), surface gravity (\logg{}), extinction (\Av{}), and veiling (\rveil{}) for young, low-mass stars, and re-evaluated the stellar properties of Tr14 using the same dataset. SAPSAL is based on a conditional invertible neural network architecture \citep[cINN;][]{ardrizzone2019b}, enabling the simultaneous estimation of these parameters from observed spectra. The network was trained on synthetic spectra generated based on the BT-Settl and BT-Dusty Phoenix stellar atmosphere libraries \citep{allard1, allard2}, in the MUSE/VLT resolution and spectral range. Their methodology enables efficient population characterization, providing stellar properties for approximately 2000 stars within minutes. Comparing their results with those of \citetalias{domi}, they found good agreement between the two classifications, and many cases where SAPSAL provided a better fit than the template fitting method. This work demonstrated the applicability of the cINN framework for characterizing young stellar populations.

A thorough investigation of the cluster core, however, where external photoevaporation is expected to be most intense, has yet to be carried out. In this paper, we present a new analysis of the stellar population in the central region of Tr14. We also present a novel tool to account for the strong background emission and extract stellar spectra from MUSE/VLT observations and we use it to acquire a robust sample from the cluster's core. Following the results of \cinn{}, we investigate the stellar properties of this previously under-characterized population with the same cINN framework and explore the impact of the local radiation environment on circumstellar disks. The paper is structured as follows: Sect.~\ref{sec:data} describes the observations and data reduction; Sect. \ref{sec:analysis} presents the stellar population analysis and parameter derivation; Sect. \ref{sec:ex_phot} describes the local FUV radiation field and examines the effect of external photoevaporation on two disk tracers; and finally, Sect. \ref{sec:summ} summarizes our conclusions. 

\section{Observations and data processing} \label{sec:data}

\subsection{Observations \label{sec:obs}}

Tr14 was observed in 2016 with the VLT/MUSE, under the program ID 097.C-0137 (PI: A. McLeod). The entire cluster was covered by 22 $1'\times1'$ pointings, with continuous spectral coverage in the range of $\mathrm{\lambda} =4650\text{-}9300$\AA, and spectral resolution varying over $R=2000\text{-}4000$ \citepalias[as described in][]{domi}. The region was observed both in short (5\,s) and long (780\,s) exposures and the data were reduced using the ESO pipeline v. 2.8.3 \citep{reduce} in the EsoRex environment \citep{esorex}, providing wavelength-calibrated IFU cubes, one per field and exposure. In this work, we analyze the four (out of the twenty-two) MUSE pointings that cover the core of Tr14, corresponding to a $2'\times2'$ region, with right ascension (RA) ranging from $156^{o}$ to $163^{o}$ and declination (DEC) from $-59.569^{o}$ to $-59.534^{o}$, and we utilize both the long and short exposures. The major limitation when this region was previously examined was, as described in \domi{}, the variable sky background, the bright and uneven halos or diffraction spikes, and the saturation trails from the bright O-type stars. In the following sections, we describe the new methodology that we developed to address these issues, and to construct the final spectroscopic catalog of Tr14's core, including spectral extraction, astrometric correction, flux calibration, and the treatment of sources observed in multiple fields. We apply our methodology for each field and exposure separately.

\subsection{Extraction of background-subtracted spectra: pySpecMUSE} \label{sec:data_extr}

For retrieving the spectra from the MUSE cubes, we present pySpecMUSE\footnote{\href{https://github.com/ltesti/pySpecMUSE}{https://github.com/ltesti/pySpecMUSE}}, a Python-based package to extract stellar spectra from MUSE data, optimized for relatively crowded fields, and bright and variable background emission. Using an implementation of the Python package \texttt{Photutils} \citep{photutils} and the DAOFIND algorithm \citep{daofind}, pySpecMUSE detects candidate stars for extraction on the MUSE \textit{I}-band synthetic images, where nebular emission is less dominant.

Given the large wavelength coverage of the MUSE spectra, the point spread function full width varies significantly across the spectrum, and consequently the flux loss within a fixed aperture as well. To correct for this effect, pySpecMUSE models the aperture correction (AC) as a function of wavelength. In our case, we use a 5th order polynomial. To compute and apply the wavelength-dependent AC, pySpecMUSE first identifies a set of bright, nearly isolated stars to serve as AC calibrators. For each of these stars, and for every layer of the MUSE cube, we apply standard aperture photometry methods to derive AC using three distinct circular apertures: a target aperture for extracting the full spectrum, a large aperture that captures the total flux from the source as defined by the growth curves, and a designated sky annulus used for sky subtraction. These three parameters can be modified by the user and need to be adjusted based on the data quality. In our case, given the plate scale, seeing, and sky background variability, we have chosen a target aperture radius of 3 pixels, a large aperture of 10 pixels, and a sky annulus from 10 to 15 pixels ($0.6^{\prime\prime}$, $2^{\prime\prime}$ and $2^{\prime\prime}-3^{\prime\prime}$ respectively). This configuration is adopted as a compromise between capturing sufficient stellar flux ($\sim 60\%$), and limiting contamination from nearby sources.

With pySpecMUSE, the sky background subtraction is made local to each star, rather than assuming that one background model represents the entire MUSE field. In addition to the background subtracted and aperture corrected spectrum, pySpecMUSE also computes the background statistics within the sky annulus and returns background spectra and background subtraction uncertainty as a function of wavelength for each target star. Using pySpecMUSE, we retrieve an initial sample of 1615 stellar spectra from the cluster's core, 905 from the long exposures and 710 from the short exposure observations.

\subsection{Astrometric alignment with Gaia} \label{sec:coords_corr}

The MUSE coordinates of each star in our sample are matched to those in the \textit{Gaia} DR3 catalog (\citeauthor{gaia1} \citeyear{gaia1}, \citeyear{gaia2}, hereafter \gaia{}). We follow an astrometric alignment procedure similar to that described by \domi{}. \gaia{} sources within each MUSE field are retrieved using the \texttt{Astroquery} package \citep{astropy2022} and iteratively cross-matched with our catalog. After each cross-match the median positional offsets in right ascension and declination are computed and applied to the MUSE coordinates using \texttt{astropy} \citep{astropy2022}. The offsets are progressively refined until the residual positional offsets converge within $0.2^{\prime\prime}$.

Table \ref{table:coords_correction} lists the final coordinate offsets applied in each field and exposure. When comparing to the similar Table B.1 of \citetalias{domi}, we note that the published offsets indicate the coordinate differences in RA, DEC, whereas in this work, we report the coordinate offsets in the plane of the sky. 

\begin{table}[ht]
    \centering
    \renewcommand{\arraystretch}{1.3} 
     \caption{Angular offsets between MUSE and \citetalias{gaia2} coordinates in each of the four fields. \label{table:coords_correction}}
    \begin{tabular}{ccc|cc}
    \hline\hline
 Field& \multicolumn{2}{c|}{Long Exposures}& \multicolumn{2}{c}{Short Exposures}\\ \hline 
         &  $\Delta \alpha$ ($''$)&  $\Delta \delta$ ($''$)&  $\Delta \alpha$($''$)&$\Delta \delta$ ($''$) \\ 
         7&  -0.72 ± 0.09&  1.54±0.07&  0.22±0.09& -2.17±0.06\\
         10&  -1.25±0.07&  1.23±0.06&  -1.00±0.08& 2.33±0.07\\
         13&  -0.98±0.08&  1.72±0.06&  0.53±0.1& -2.43±0.07\\
         16& -0.88±0.08& 2.92±0.06& -0.56±0.09& 2.17±0.07\\ \hline
    \end{tabular}
   
\end{table}

\subsection{Wavelength-dependent flux calibration} \label{sec:lambda_corr}

While the MUSE pipeline performs a basic flux calibration of the cubes, this is not sufficiently accurate for our purposes. We therefore calibrate our spectrophotometry to the \citet{wfi} Wide Field Imager (WFI) catalog. Since the observations were not performed simultaneously, we use an average flux correction. The procedure is similar to that in \domi{}, with the important improvement that we use all three WFI magnitudes (from the \textit{V, R} , and \textit{I} filters) to perform a linear fit for each of the MUSE fields. With this method, we provide a wavelength-dependent photometric correction for the spectra. We note, however, that we are aware of the high variability of young stars. Therefore, because the WFI catalog is not contemporaneous with our observations, and given the uncertainty in the absolute flux introduced by the intrinsic variability of classical T-Tauri stars \citep[CTTS, typically about 0.5 mag;][]{lorenzetti2013}, this translates in an uncertainty of about 50\% on the absolute flux calibration of our spectra.
 In Table \ref{table:phot_corrections} we present our results for each field and exposure separately.

\begin{table}[ht]
\centering
\renewcommand{\arraystretch}{1.1} 
\caption{Photometric corrections for our fields, computed as the median differences between MUSE and WFI \citep{wfi} magnitudes.\label{table:phot_corrections}}
\begin{tabular}{c|ccc}
\hline\hline
Field & \multicolumn{3}{c}{Long Exposures} \\ 
 & \begin{tabular}[c]{@{}c@{}}I\\(mag)\end{tabular} & \begin{tabular}[c]{@{}c@{}}R\\(mag)\end{tabular} & \begin{tabular}[c]{@{}c@{}}V\\(mag)\end{tabular} \\
7 & -0.15±0.10& -0.14±0.16& -0.16±0.30\\
10 & -0.10±0.19& -0.08±0.22& -0.11±0.25\\
13 & -0.07±0.22& -0.05±0.26& -0.07±0.36\\
16 & -0.12±0.11& -0.08±0.15& -0.10±0.27 \\ 
\end{tabular}

\centering
\renewcommand{\arraystretch}{1.2} 
\begin{tabular}{c|ccc}
\hline\hline 
Field 
& \multicolumn{3}{c}{Short Exposures}  \\ 
 & \begin{tabular}[c]{@{}c@{}}I\\(mag)\end{tabular} & \begin{tabular}[c]{@{}c@{}}R\\(mag)\end{tabular} & \begin{tabular}[c]{@{}c@{}}V\\(mag)\end{tabular} \\ 
7 & -0.11±0.10& -0.17±0.16& -0.24±0.38\\
10 & -0.10±0.20& -0.08±0.22& -0.11±0.25\\
13 & -0.07±0.22& -0.05±0.26& -0.07±0.36\\
16 & -0.13±0.16& -0.17±0.19& -0.29±0.43 \\ 
\end{tabular}

\end{table}

\subsection{Final spectral catalog \label{sec:final_catal}}

Before finalizing our spectral catalog, we account for duplicate stars between different fields and exposures. The MUSE pointings overlap slightly at the boundaries of adjacent fields. We identify 32 common stars between the long exposure pointings and 6 common stars between the short exposure pointings. In addition, we identify 665 common sources between the two exposures. For each pair of common stars, we discard the spectrum with the lower signal-to-noise ratio (SNR). We also remove, based on visual inspection, spectra that show strong background contamination or poor extraction quality.

Finally, we evaluate the SNR of each spectrum at the MUSE \textit{I} and \textit{V} bands. From our initial catalog, we retain all spectra of sufficient quality ($\mathrm{SNR_{\textit{I}}}>3$ and  $\mathrm{SNR_{\textit{V}}}>1$). The signal to noise thresholds are set at the minimum value that ensures adequate retrieval of the physical parameters, based on the analysis performed by \cinn{}. Our final spectral sample from the center of Tr14 consists of 421 targets, 392 from the long exposures and 29 from the short exposure observations. This is a significant increase in the number of stars detected in the center of the cluster compared to the results of \citetalias{domi}. In their study, they identify 780 stars with robust photometry, but only 150 spectra in the center of the cluster. Our improved methodology for accounting for the variable sky background, enables the extraction of higher-quality spectra, increasing this central sample by 2.8 times.

\subsection{NIR photometry}

To further characterize the properties of our sample, we use NIR photometry (\textit{J}, \textit{H}, and \textit{K} bands) from the HAWK-I \citep{preibis1, preibis2} and VISTA \citep{preibis3} surveys of the CNC. Before cross-matching with our sample, we align the astrometry of the two surveys with \citetalias{gaia2}, following the same methodology as for the astrometric correction of the MUSE coordinates. We match each NIR table with \citetalias{gaia2} adopting $0.5''$ as the maximum separation limit. We deduct the coordinate offsets as listed in Table~\ref{table:nir_correction}.

\begin{table}
    \centering
    \renewcommand{\arraystretch}{1.3} 
\caption{Angular offsets between HAWK-I and VISTA surveys, using the \citetalias{gaia2} catalog}
\label{table:nir_correction}
    \begin{tabular}{c|cc}
    \hline\hline
        & $\Delta \alpha$ ($''$) & $\Delta \delta$ ($''$)\\ \hline
        HAWK-I & $-0.11 \pm 0.06$& $-0.11 \pm 0.06$\\ 
        VISTA  & $-0.16 \pm 0.06$& $-0.002 \pm 0.055$\\ 
    \hline
    \end{tabular}
    
\end{table}

We select the HAWK-I survey as our main reference for the NIR photometry, due to its capability to detect faint stars~\citep[$J\approx23$ mag,][]{preibis1}. For comparison, the VISTA survey identifies stars with magnitudes down to $J\approx21.5$ mag \citep{preibis3}. In total, we identify counterparts for 391 stars of our sample: 367 stars matched with the HAWK-I survey and 24 stars matched with VISTA. The unmatched stars are due to the incompleteness of the HAWK-I and VISTA source catalogs in the very dense and crowded core of the cluster. By obtaining the NIR photometry, we are able to calculate the bolometric luminosity of the sources (see Sec. \ref{sec:s_params}) and identify NIR excess sources (see Sec. \ref{sec:nirex}).

\section{Stellar population analysis \label{sec:analysis}}

\subsection{Application of SAPSAL in the core of Tr14} \label{sec:sapsal_on_tr14}

In \cinn{}, \texttt{SAPSAL-v2} was applied to the spectral sample acquired in \domi{}, deriving the stellar properties from the entire region of Tr14. In this work, we concentrate on the center of the cluster, and we apply this version of SAPSAL to the increased spectral sample described in Sec.~\ref{sec:final_catal}. To prepare our spectra for SAPSAL, we first mask out the spectral regions excluded from the training process, such as emission lines, and then normalize the spectra by dividing them by the total flux of the remaining spectral bins. We also determine the relative flux error following the method in \cinn{}, by calculating the standard deviation of the flux within the central region of the MUSE spectra (i.e., $705\pm40\,\mathrm{nm}$), which range from 470 to 940~nm. To align with the SAPSAL training, which assumes a wavelength-independent error, we apply the error determined in this central region uniformly across the entire wavelength range. Applying SAPSAL to the normalized spectra, we obtain a multi-dimensional posterior distribution and adopt the maximum a posteriori (MAP) values from the one-dimensional posterior distribution of each parameter as the representative estimates for \teff{}, \logg{}, \Av{}, and \rveil{}. The MAP values are determined by fitting Gaussian kernel density estimations on the 1D posterior distributions. The MAP estimate corresponds to the point where the probability density reaches its maximum.

From the results, we identify a subset of stars with parameter estimates outside the training limits of SAPSAL. Specifically, SAPSAL is trained on spectra spanning a grid in \teff{} ranging from $2600$ to $7000\, \mathrm{K}$. Moreover, to account for the observational properties of the spectra, \citetalias{cinn2} applied veiling as a constant continuum excess, and extinction following the extinction law by \cite{cardeli} with $R_{V}=4.4$ \citep{hur}, specifically tailored for the Tr14 cluster. We identify 68 stars with $T_{\mathrm{eff}}^{\mathrm{cINN}}>7000\mathrm{K}$ and 47 with $A_{\mathrm{V}}<0$. Extrapolated outputs in our sample can occur due to contamination either from surrounding sources or initially in the data reduction process. When a spectrum falls outside the training domain of SAPSAL, it prevents the network from inferring valid physical parameters for the sources. To handle the sources with \teff{} exceeding the training limits, we completely exclude them from our further analysis. Concerning the stars with $A_{\mathrm{V}}<0$, we follow the approach by \citetalias{cinn2}: we only keep the stars with predictions that fall within a $\pm 0.1\mathrm{mag}$ margin from the boundaries of \Av{}. In this process, we discard the 47 stars with $T_{\mathrm{eff}}^{\mathrm{cINN}}>7000\mathrm{K}$, and 21 stars outside the \Av{} margin, resulting in a final sample of 353 stars with parameter estimates within the SAPSAL training range.

\subsection{Stellar parameters } \label{sec:s_params}

We examine the stellar population in our sample by constructing the Hertzsprung–Russell (HR) diagram. To calculate the stellar luminosities, we first determine the spectral type (SpT) and bolometric correction (BC) of each star using the SAPSAL-estimated \teff{} and the \teff--SpT scales from \cite{kh} (for types B0 to G9 , $T_{\mathrm{eff}}= 5410\text{K} - 30000 \text{K}$) and \cite{lhu} (for types K0 to M9, $T_{\mathrm{eff}}= 2400\text{K} - 5250 \text{K}$). For cases in between the reported scales, we extract SpT and BC by linear interpolation.

We follow the approach of \domi{} considering \textit{J}-band photometry \citep[taken from][]{preibis1,preibis2, preibis3}, as it minimizes the contamination of mass accretion or intrinsic differential extinction on the stellar spectrum, and we use the following relations to compute $L_{\text{bol}}$:
\begin{equation}
    M_{\mathrm{bol}} = J - A_{J} - DM + (BC_{\mathrm{V}} + (V - K) - (H - K) - (J - H)),
\end{equation}
and
\begin{equation}
    \log(L_{\mathrm{bol}}/L_{\odot}) = -0.4 \cdot (M_{\text{bol}} - M_{\text{bol},\odot}),
\end{equation}
where we adopt $11.86 \text{ mag}$  for the distance modulus \citep[DM,][]{dm}, and $4.74\ \text{mag}$ for the solar bolometric magnitude \citep[$M_{\mathrm{bol,\odot}}$,][]{mbol}. To compute the extinction in the \textit{J}-band ($A_J$), we follow the extinction law by \citet{cardeli}, with $R_{\mathrm{V}} = 4.4$ \citep{hur}. Under the assumption that main-sequence stars do not exhibit accretion, and given that the bolometric luminosities are derived from the \textit{J}-band, we can adopt $L_{\mathrm{star}} \approx L_{\mathrm{bol}}$ for describing the entire sample. The resulting luminosities therefore primarily trace the stellar photosphere.

\begin{figure}[h!]
    \centering
    \includegraphics[scale = 0.38]{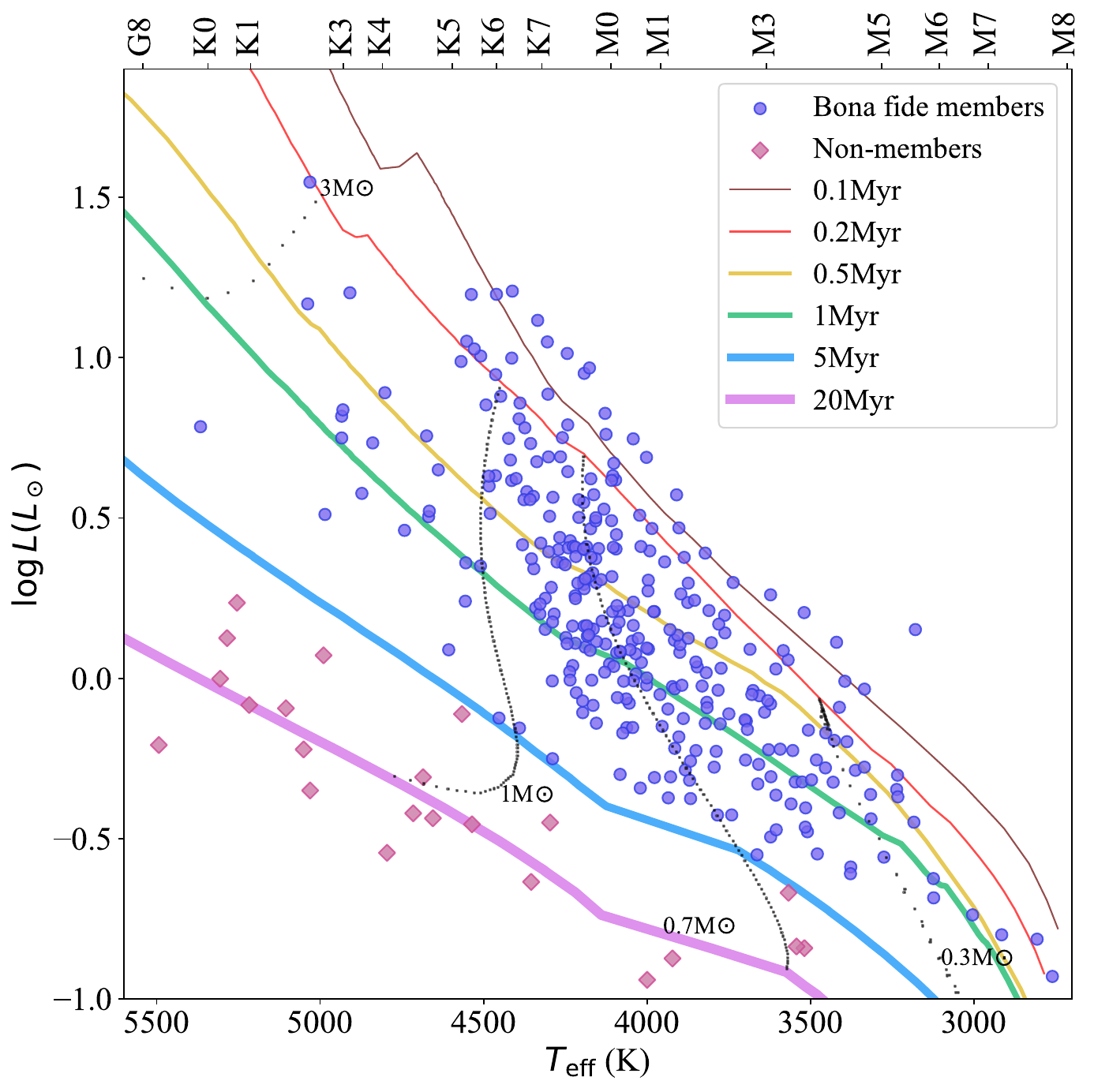}
    \caption{HR diagram of our stellar sample of the Tr14 center. Blue circles represent the bona fide cluster members, whereas pink rhombuses represent stars considered as non-members. The evolutionary tracks from \citet{parsec} are shown as colorful lines indicating the isochrones; dotted gray lines indicate the isomasses.}
    \label{fig:HR}
\end{figure}

In Fig. \ref{fig:HR} we present our sample on the HR diagram. The isochrones and isomasses displayed are computed using the PARSEC evolutionary tracks \citep{parsec}. We have restricted our final sample to stars with $T_{\mathrm{eff}} \leq 5500$ K, where the HR diagram shows a coherent stellar sequence well described by the adopted evolutionary tracks (see Appendix ~\ref{app:weird_s}).

We derive the age and mass of each star by performing linear interpolation between the PARSEC tracks.  Using the HR diagram as a reference, we examine the spectral-type distribution of the young population (<5 Myr). Most objects in this subsample lie between K4 and M3, though we also identify a few stars as late as M8 and as early as G8. In Fig. \ref{fig:HR}, we observe that most of the population is concentrated near the $1\text{ Myr}$ isochrone and extends towards younger ages. In addition, there is a subpopulation situated close to the 20 Myr isochrone, with some sources approaching 5 Myr. Considering previous estimations of the cluster's young age \citep[e.g.,][\domi{}]{preibis1}, we investigate the spread in the age distribution of our sample. We find that the distribution peaks at $0.83\ \text{Myr}$ and shows an extended tail towards older ages up to $63$ Myr. The Gaussian fit to the distribution yields a dispersion of $\sigma=0.32$ dex, placing the 5 Myr threshold at $+1.61\sigma$ from the peak. Thus, while the bulk of the sources is associated with the young population, the older tail introduces a contribution from significantly older objects. To ensure that our analysis focuses on the population linked to recent star formation, we restrict the sample to stars younger than 5 Myr. Sources above this limit are therefore considered part of the older tail and are excluded from the subsequent analysis. We note that when spatially examining this older population, we find that it is homogeneously distributed throughout the region.

\begin{figure}[h]
    \centering
    \includegraphics[width=0.48\textwidth]{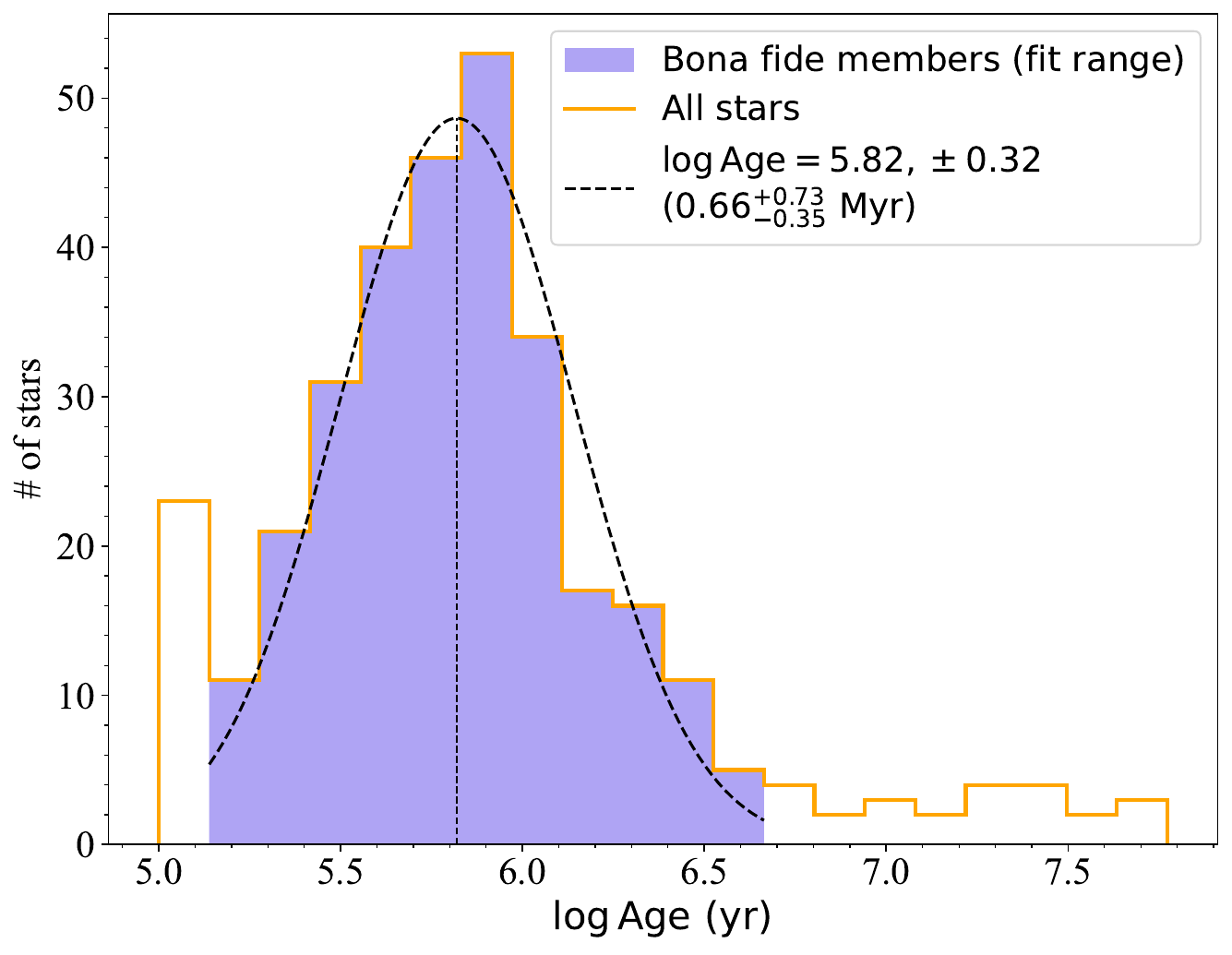}
    \caption{Histogram of the age distribution of bona fide members in the central region of Tr14.\label{fig:age}}

\end{figure}

We identify the stellar sample younger than 5 Myr threshold as bona fide members and we display their age distribution as a purple-filled histogram in Fig. \ref{fig:age}. The bona fide members consist of 310 stars. The outlined orange histogram indicates the age distribution of the total sample, including the non-members. To calculate the age of the cluster's center, we fit a log-normal function to the histogram. For the sources situated beyond the boundaries of the PARSEC theoretical models, their ages get fixed at the upper or lower limit accordingly, causing an over density in those bins. We therefore omit the first and last bins from the calculation of the average age. From the bona fide members, we compute a cluster age of $0.66 ^{ +0.73}_{-0.35} \text{  Myr}$, consistent with previous estimates \citepalias[e.g.,][]{domi, cinn2}. We further examine the age distribution as a function of projected distance from the cluster center, combining the central sample analyzed in this work with the outer-cluster sample from \cinn{}. Within the uncertainties of the age estimates, we find no evidence for a systematic spatial trend in age, either in the cluster center or across Tr14 as a whole.

\begin{figure}[h]
    \centering
    \includegraphics[width=0.42\textwidth]{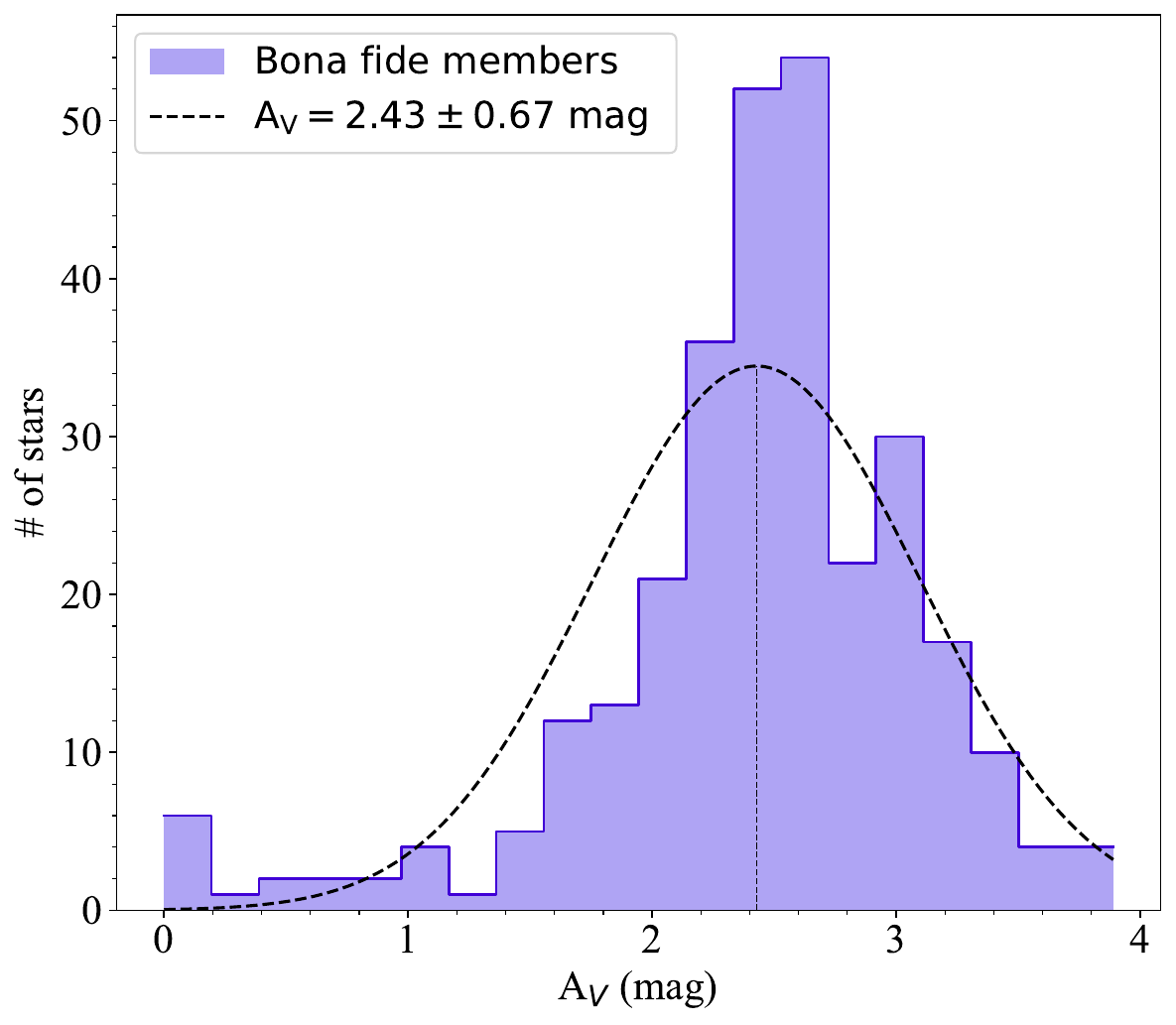}
    \caption{Histogram of the interstellar extinction distribution in the central region of Tr14. \label{fig:av}}
\end{figure}

Additionally, we examine the distribution of the visual extinction towards the region in Fig. \ref{fig:av}. We calculate an extinction of $A_{\mathrm{V}} = 2.43\pm0.67\text{ mag}$ by fitting a Gaussian function to the distribution. We therefore find that, after accounting for the relevant uncertainties, our result agrees with earlier findings reported in the literature \citep[][\domi{}]{ascenso}.

Finally, we investigate the mass distribution of our sample obtained through the PARSEC isomasses. The stellar sample from the center of Tr14  has masses ranging from $0.09\ \mathrm{ M_{\odot}}$ to $3.20\ \mathrm{M_{\odot}}$. The distribution appears to follow the shape of the initial mass functions (IMFs) of \citet{kroupa} and \citet{chabrier} for $m_{*}>0.5\ \mathrm{M_{\odot}}$, showing a sharp decline for $m_{*}>0.9\ \mathrm{M_{\odot}}$. For stellar masses below $0.5\ \mathrm{M_{\odot}}$, however, the distribution falls off relative to the IMFs, most notably for $m_{*}<0.2\ \mathrm{M_{\odot}}$. We interpret this as the result of our incompleteness arising from the difficulty of detecting faint, low-mass stars with sufficiently high SNR.

\subsection{NIR excess classification \label{sec:nirex}}

\begin{figure}[h!]
    \centering
    \includegraphics[scale = 0.4]{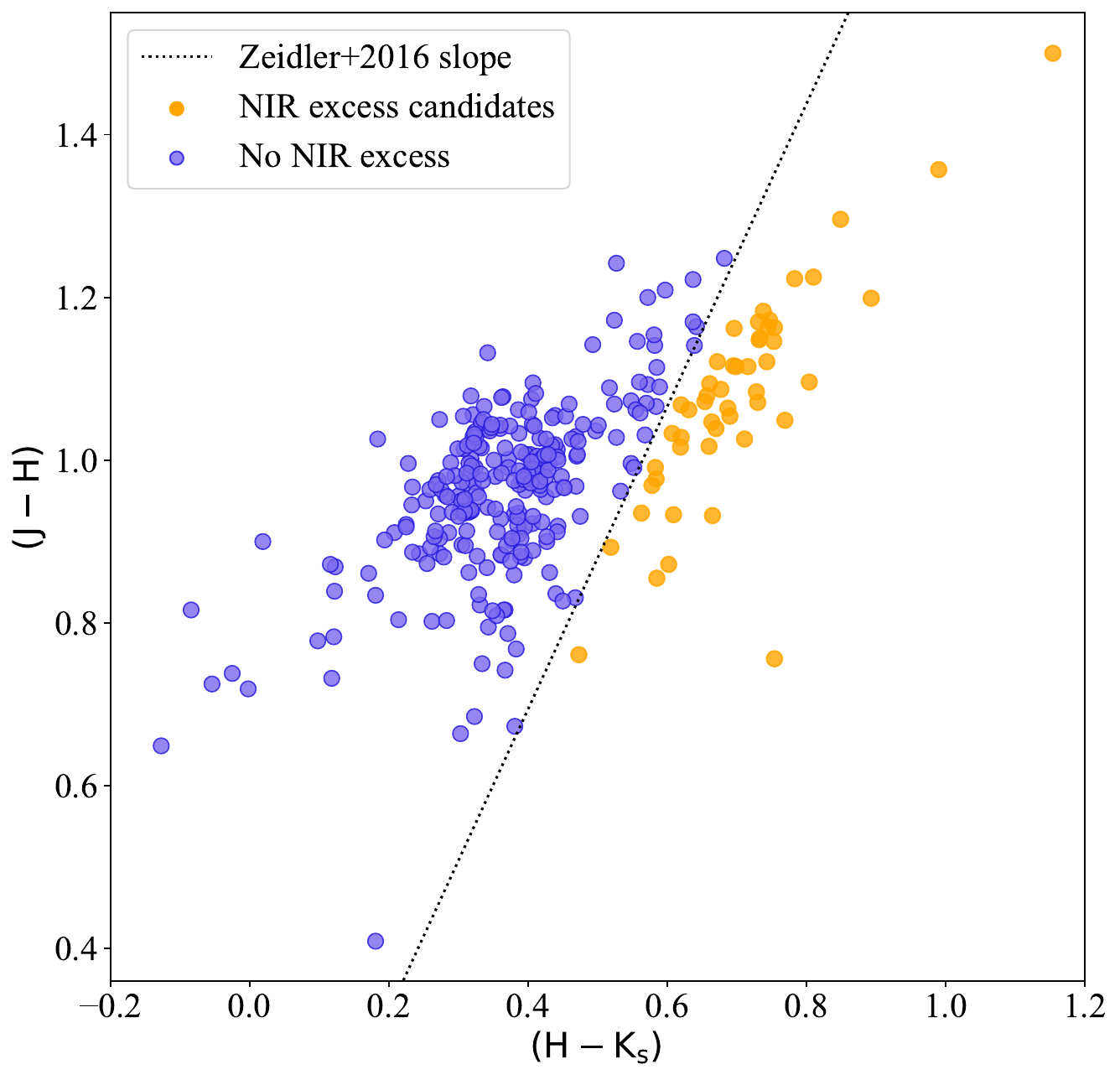}
    \caption{(${J-H}$) vs. (${H-K_s}$) color-color diagram of our bona fide members. The dashed line represents the empirical law found by \citet{nir} , describing the slope of the reddening vector in the CNC. The line corresponds to a slope of 1.86. All stars found with $J-H$ > 0.05 mag, $H-K_s$ > 0.05 mag and more than 0.05 mag bellow and to the right of the slope are identified as NIR excess sources, and are marked with orange circles. The average photometric errors reported in the NIR catalogs are almost negligible, and therefore not displayed in the figure.\label{fig:NIR_excess_zeidler} }
\end{figure}

Using the NIR photometry acquired from the \citet{preibis1, preibis2, preibis3} catalogs, and the approach presented by \citet{nir}, we identify the sources with NIR excess when they are placed 0.05 mag below and on the right of a 1.86 slope in the $(J-H)$ vs. $(H-K_s)$ diagram. For classifying the stars, this approach considers the position of the color-color measurements, without their uncertainties. We display the $(J-H)$ vs. $(H-K_s)$ diagram in Fig. \ref{fig:NIR_excess_zeidler}. The photometric errors reported for the HAWK-I and VISTA catalogs (which are the statistical errors reported by the original photometry routine) are very small for most sources in our sample, therefore they are not taken into consideration for the selection of the sources. We examine the photometric uncertainties and their impact on this selection in Appendix ~\ref{app:nir_err}. Following this methodology, we identify 17\% of our sample (52 stars) as NIR excess sources, slightly higher than the 9.7\% disk fraction found by \citet{preibis1}. This difference likely arises because they analyzed an X-ray–selected stellar sample spanning the full extent of the Tr14 cluster, whereas our study focuses on an age-selected sample within its central region. In addition, their analysis adopted a slightly different extinction law, whereas our analysis employs the most recent determination for the CNC \citep{nir}.

\begin{figure}[h!]
    \centering
    \includegraphics[width=0.5\textwidth]{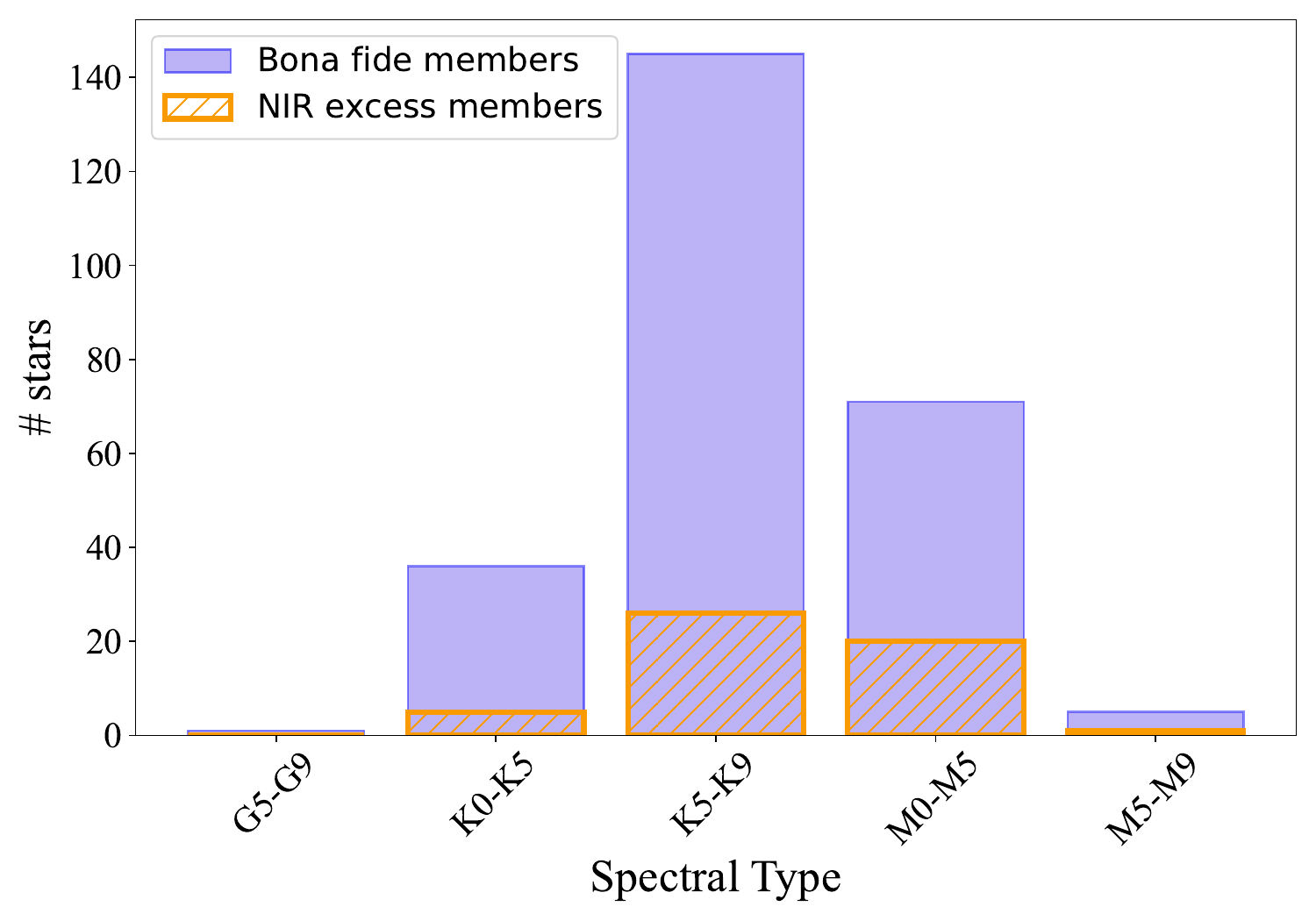}
    \caption{Comparison of the spectral-type distribution of the bona fide members in our spectral sample (in purple, filled bars) with the subsample showing NIR excess (orange, hatched bars).
    \label{fig:spt_hist_wth_nir} }
\end{figure}

We further investigate the selected NIR excess sources by examining their spectral types and comparing their distribution to that of the total sample. The comparison is displayed in Fig. \ref{fig:spt_hist_wth_nir}. We divide each spectral class in early and late type and we find that the distribution of the NIR excess sources follows the shape of the total sample.

\subsection{Comparison with previous studies \label{sec:prev_stud}}

\begin{figure}[h]
    \centering
    \includegraphics[width=0.45\textwidth]{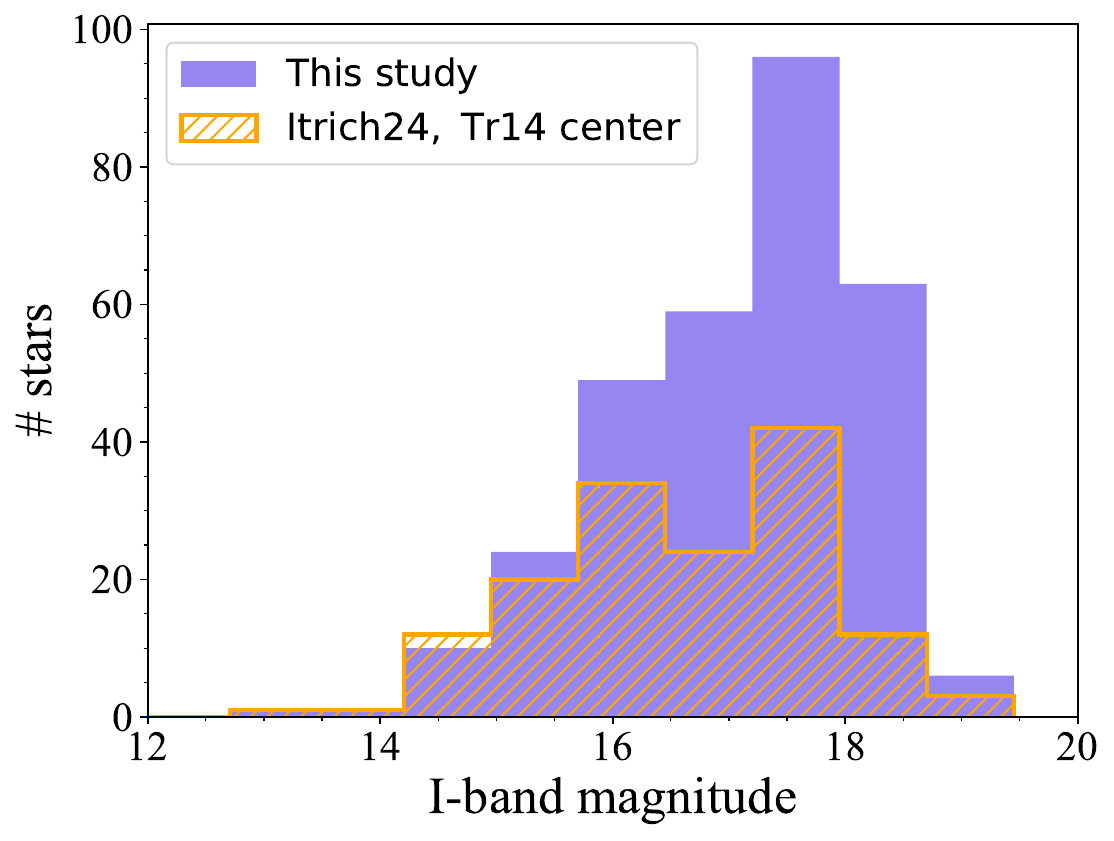}
    \caption{The distribution of our sample in \textit{I}-band magnitudes, in comparison to the central sample of \domi{}. As we go to fainter magnitudes, we see that pySpecMUSE is able to retrieve more spectra with reliable photometry. \label{fig:i-band_comp}}
\end{figure}

In this paper we have re-extracted and re-analyzed the central regions of the Tr14 MUSE/VLT data described by \domi{} and \cinn{}. Compared to these studies, and as described in the previous sections, we have used a new and more accurate methodology to estimate and remove the variable sky background before extracting the stellar spectra. In addition, we have employed a wavelength-dependent approach for the aperture correction and flux calibration. The use of pySpecMUSE has allowed us to reliably extract spectra for 421 stars from the inner regions of the cluster, as compared to the 170 stars extracted by \domi{} in the same region. If we account for our sample cuts, we characterize 332 targets, 2.2 times more than the 150 targets with high photometric certainty by \domi{}. In Fig. \ref{fig:i-band_comp} we compare the last two samples as a function of their \textit{I}-band magnitude. With pySpecMUSE, we are able to retrieve two to six times more spectra for the faint end of the magnitude distribution (\textit{I}>16). This improvement is mainly due to a more accurate estimate and subtraction of the variable sky background, and the use of procedures optimized for stellar photometry in crowded fields.

\begin{figure}[h]
    \centering
    \includegraphics[width=0.5\textwidth]{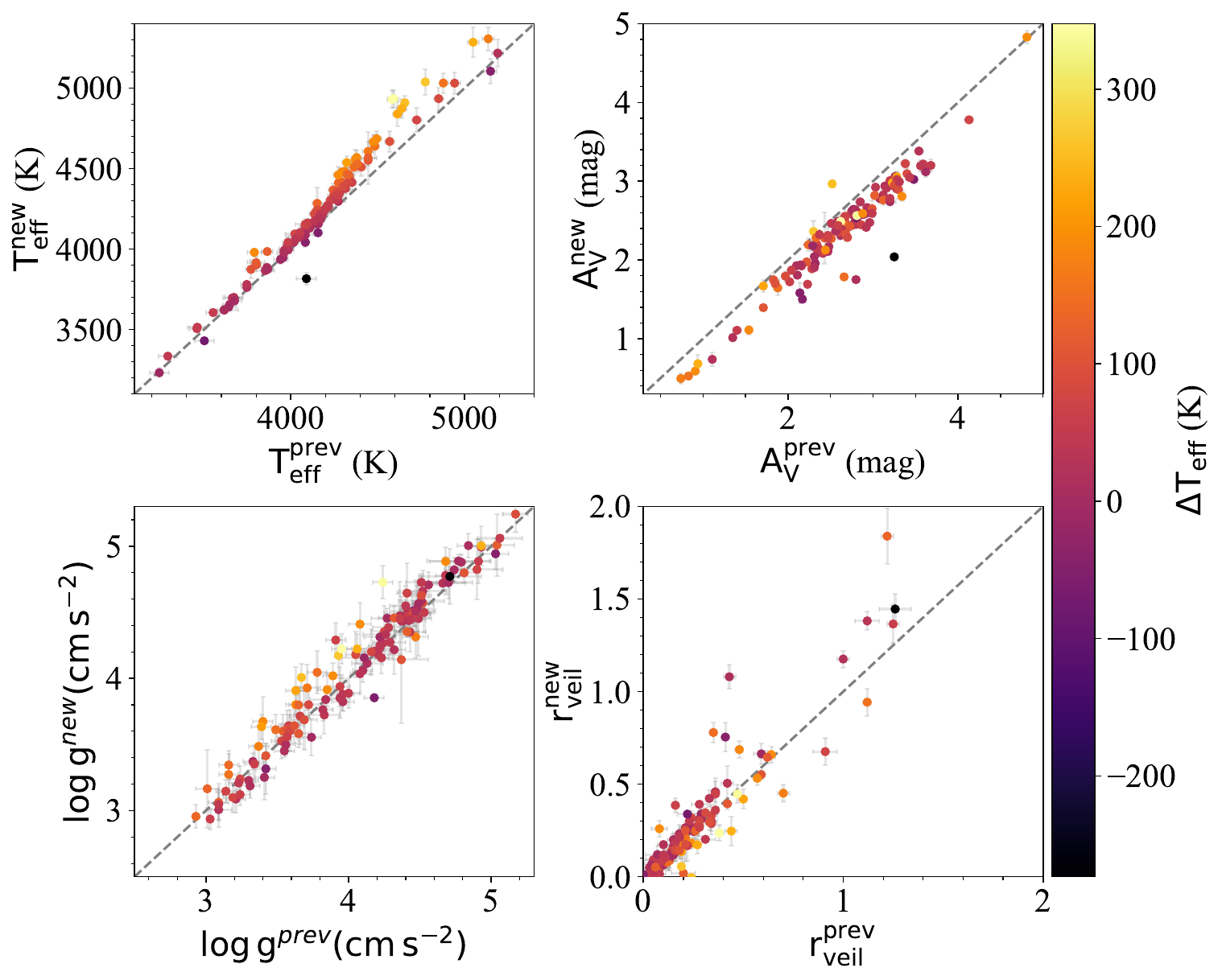}
    \caption{Comparison of the estimated parameters $T_{\text{eff}}$, \logg{}, \Av{} and \rveil{} from our current analysis ("new") with those by \citetalias{cinn2} ("prev") for the 129 common stars. The color-bar in all 4 figures shows the difference in temperature estimations between the two works. \label{fig:param_comp_kang}}
\end{figure}

Furthermore, we examine the impact of the wavelength-dependent aperture correction and flux calibration on the cINN results. In Fig. \ref{fig:param_comp_kang} we compare the four cINN-estimated stellar parameters of the stars in common with \cinn{}. Our final sample only has 129 stars in common with the central sample by \domi{}, as we have excluded many spectra due to their SNR or due to the strong background contamination. The x-axis of each panel indicates the parameters estimated by \cinn{} using the spectra from \domi{} and denoted as \textit{"prev"} (previous). The y-axis indicates the parameters estimated in this study, denoted as \textit{"new"}. The color indicates the MAP temperature difference between this study and \cinn{}. Compared with the simpler, non-wavelength-dependent corrections of \domi{}, our more accurate methodology produces spectra with a different slope, generally bluer. The main effect is in a systematic reduction of the measured extinction values (by $0.28\ \text{mag}$ on average), and, to a lesser extent, a modification of \rveil{} and \teff{}. The latter affects mostly stars hotter than $4500\ \text{K}$.   

\begin{figure}[h]
    \centering
    \includegraphics[width=0.43\textwidth]{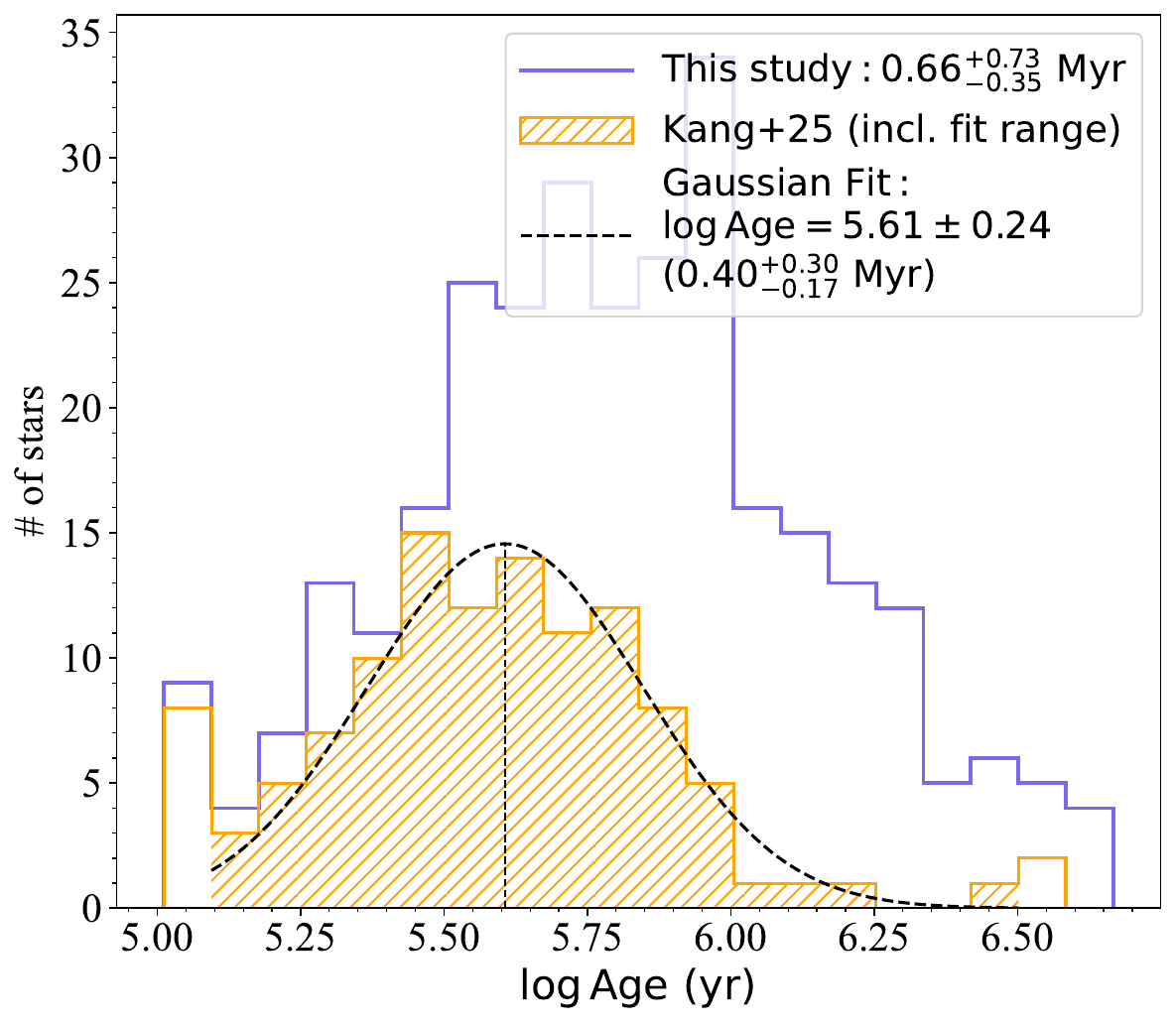}
    
    \caption{Comparison of the age distribution derived in this study with the one from  \citetalias{cinn2}, considering only the central region of Tr14. The ages are estimated based on a Gaussian distribution fitted on the data (black dashed curve). For the average age, we only use the hatched distribution which excludes the first and last bins, to remove contamination from the PARSEC boundaries.   \label{fig:age_av_comp_kang}}
\end{figure}

Finally, we compare the average ages of the stars in the central region of Tr14 as derived from the template fitting method \citepalias[$0.66^{+0.68}_{-0.34}\ \mathrm{Myr}$,][]{domi}, the application of SAPSAL on those spectra by \cinn{} ($0.40^{+0.30}_{-0.17}\ \mathrm{Myr}$, seen in Fig.~\ref{fig:age_av_comp_kang}), and our own spectra analyzed with SAPSAL. The three values are consistent within the relatively large uncertainties.

\section{External photoevaporation of Tr14's disks \label{sec:ex_phot}}

In this section we will use our extensive spectral catalog and broad band photometry to investigate the effect of external photoevaporation on the disks surrounding low mass stars in the center of the cluster. To this aim we first derive an estimate of the local external FUV radiation for each one of the young stars, and then we explore the correlation of the flux estimates with the derived optical veiling and the fraction of sources with infrared excess.

\subsection{The local FUV field \label{sec:fuv}}

To estimate the FUV flux, we employ the method presented by \cite{g0} and we obtain the incident radiation flux in Habing flux units \citep[$G_\mathrm{{0}} = 1.6\times10^{-3}\text{erg } \mathrm{s^{-1}cm^{-2}}$, ][]{habing} for each star in our sample. The geometric FUV flux experienced by each target star $i$ is computed by summing the radiation contributions from all massive stars ($N_{\mathrm{OB}}$) in the region:
\begin{equation}
F_{\mathrm{FUV}, i} = \sum_{j=1}^{N_{\mathrm{OB}}} \frac{L_{\mathrm{FUV}, j}}{4 \pi d_{ij}^{2}},
\end{equation}
where $L_{\mathrm{FUV}, j}$ is the FUV luminosity of the $j$-the OB star, and $d_{ij} = |\mathbf{x}_{i} - \mathbf{x}_{j}|$ represents their true 3D spatial separation.

\begin{figure}[h!]
    \centering
    \includegraphics[width=0.42\textwidth]{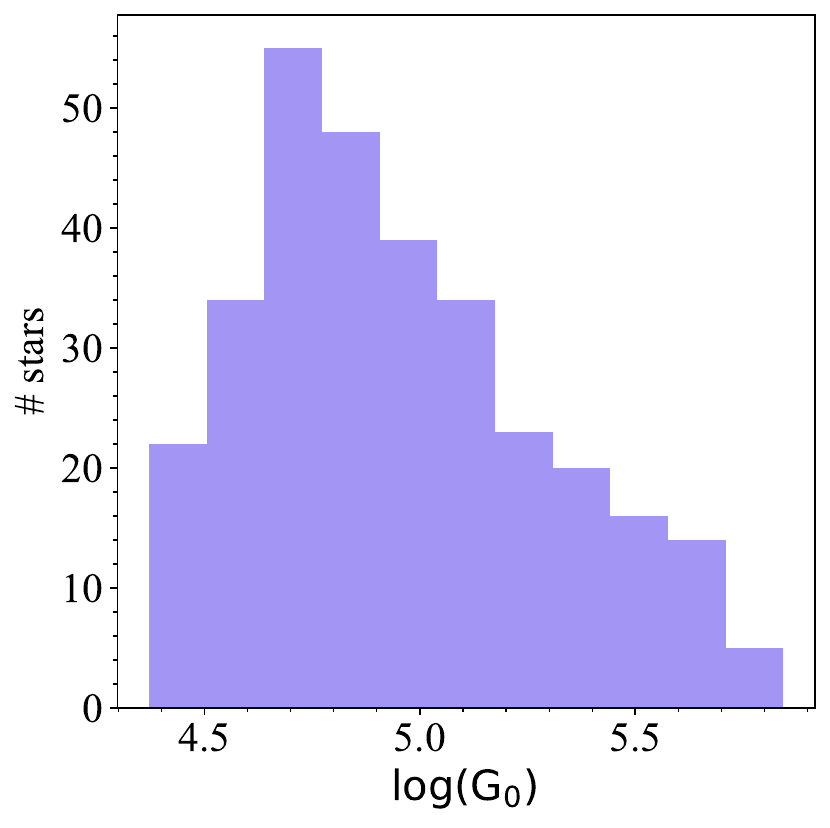}
    \caption{Distribution of the median FUV flux estimated by assuming a centrally concentrated Tr14 core, and considering the 3D separations of the low-mass stars and their OB-type neighbors. The FUV flux is displayed in logarithmic scale, and in units of \go{}. \label{fig:g0_dist}}
    
\end{figure}

To account for the large observational uncertainties in the line-of-sight positions of these stars, and consequently in their 3D separations, we adopt a probabilistic approach using the observed 2D projected geometry of the stellar cluster to infer its 3D stellar volume density distribution \citep{g0}. Tr14 is highly centrally concentrated around its most massive O-type star, with a core radius of $0.8\text{ pc}$ \citep{ascenso}. To compute the 3D separation, we adopt this approach, approximating the cluster as spherically symmetric to the first order. By mapping the 2D projected radial distances ($R$) of the member stars from the cluster center, the 3D volume density profile $\rho(r)$ is derived via an Abel inversion of the observed surface density profile. For Tr14, this yields a spatial volume density distribution of the form $\rho(r) \propto \left(1 + (r / 0.8\,\text{pc})^2\right)^{-4}$, which, once normalized, provides the probability distribution of the 3D distance $r$ from the cluster center \citep{ascenso, g0}. By executing a Monte Carlo sampling of the resulting posterior distribution of 3D separations from the OB stars in the cluster's center, we compute the best-estimate FUV fluxes and their associated geometric uncertainties, defined by the 16\%, 50\% (median), and 84\% percentiles. The computed FUV flux ranges from approximately $10^{4.2}$ \go{} to $10^{6.0}$ \go{}, considering the uncertainties, and the distribution peaks at $\sim 10^{4.7}\ {G_{\mathrm0}}$ (Fig. \ref{fig:g0_dist}). Finally, we compare the FUV fluxes derived using projected 2D separations with those inferred from the probabilistic 3D geometry. Accounting for the cluster's 3D structure systematically lowers the estimated FUV fluxes: by approximately 20-40\% for $\log(F_{\rm FUV}) > 5$, and by less than 20\% for $\log(F_{\rm FUV}) < 4.75$.

\subsection{Dependence of veiling to the local FUV field \label{sec:veiling}}

Optical veiling in SAPSAL-v2 of \cinn{}, is modeled as a constant excess flux at all wavelengths. This is, of course, a first order approximation, which is aimed to estimate the effect of the excess emission around the YSO, on the spectra. In the optical regime, excess emission is typically linked to the accretion shock onto the central star \citep[e.g.,][]{veil_calvet, veil_herczeg}, and is observed as a weakening of the photospheric absorption features of the stellar spectrum \citep[e.g.,][]{veil_basri,veil_manara}. Because the amount of veiling is directly tied to the accretion flow onto the star, it serves as an indirect tracer of the presence and activity of the circumstellar disk. Given that intense far-ultraviolet radiation fields can significantly modify the gaseous content of circumstellar disks \citep[e.g.,][]{clarke2007, concha2019, coleman2022, wh}, an indirect effect on accretion-related observables such as veiling is expected. Specifically, a stronger FUV field should deplete the gas reservoir, thereby lowering accretion rates and, consequently, reducing the observed veiling factors.

\begin{figure}[h!]
    \centering
    \includegraphics[scale = 0.4]{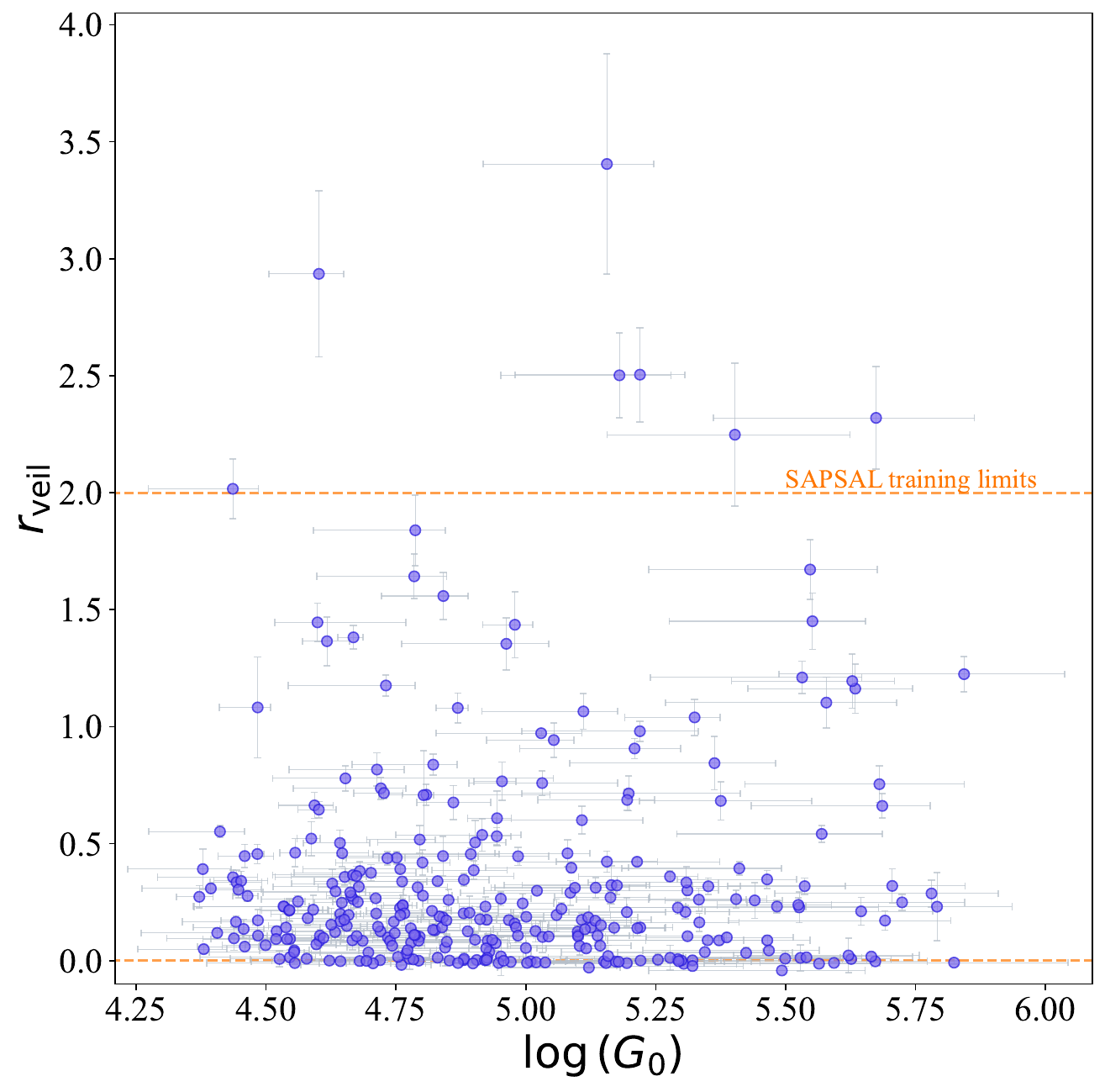}
    \caption{Distribution of \rveil{} as a function of the FUV field in units of \go{}. The FUV flux is displayed in the logarithmic scale and the displayed sample represents the bona fide members. With orange dashed lines, we indicate the SAPSAL training limits for \rveil{} (0-2). \label{fig:veil_g0} }
\end{figure}

\begin{figure}[h!]
    \centering
    \includegraphics[scale = 0.45]{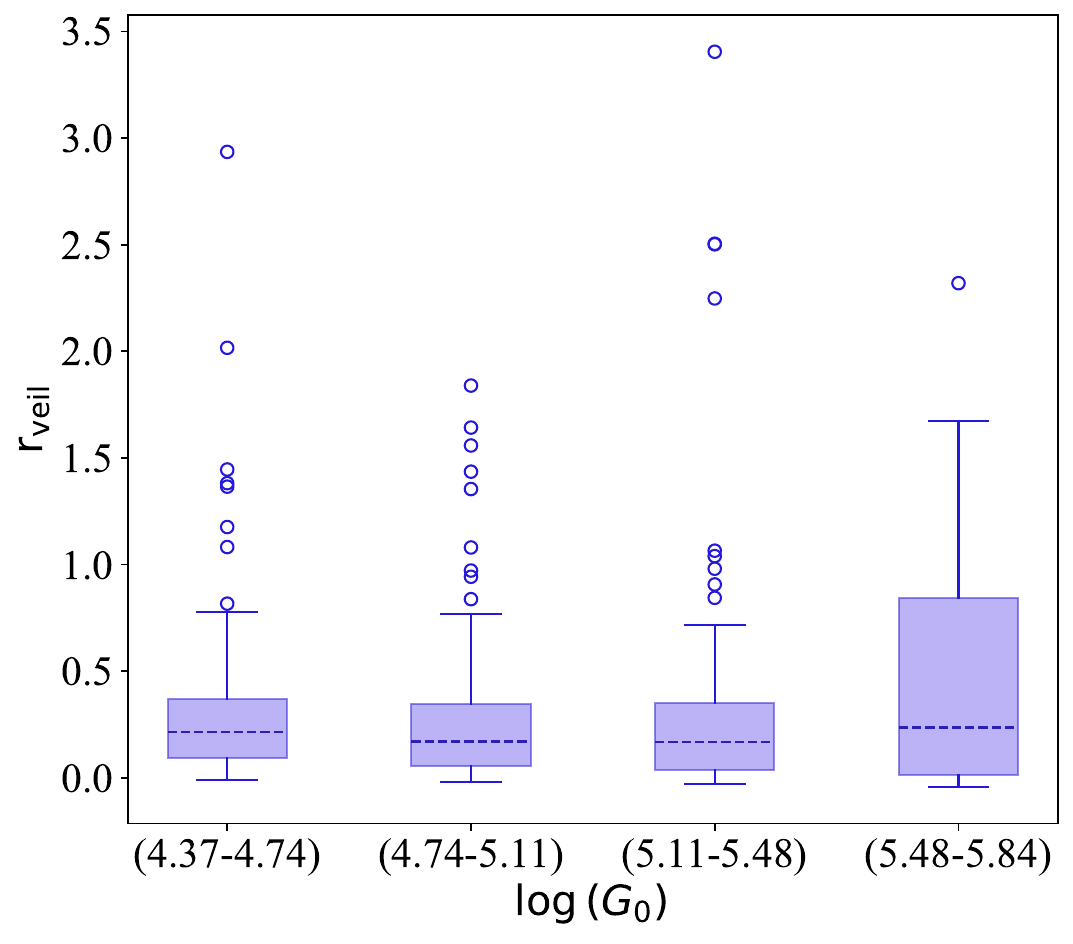}
    \caption{Box plot of the distribution of \rveil{} in function of binned \go{}. The FUV flux is displayed binned within the estimated total range of \go{} in 4 groups and in the logarithmic scale. The dashed line indicates the mean value of \rveil{} in each bin. \label{fig:binned_veil_g0} }
\end{figure}

Figure~\ref{fig:veil_g0} shows the distribution of spectroscopic veiling (\rveil{}) as a function of the local FUV radiation field in units of \go{} for our stellar sample. We find that 23\% of the sample has $r_{\text{veil}}\approx0$, while the majority of sources are concentrated at low veiling values ($0<r_{\text{veil}}<0.25$), independent of the local FUV field. In addition, 35\% of the sample exhibits relatively high veiling values, with $r_{\text{veil}}>>0.3$. We note that in our study, we include the sources with \rveil{} surpassing the SAPSAL training limits ($r_{\text{veil}}>2$), as we expect highly veiled spectra based on the age distribution of the members. To compare the variation of veiling based on the strength of the FUV field, we calculate the distribution of veiling in binned sections of the FUV flux (Fig. \ref{fig:binned_veil_g0}). We divide the FUV flux in 4 bins, to retain a similar amount of stars per bin. We infer a median value for the veiling parameter of $r_{\text{veil}}^{med}\approx 0.2\pm0.04$ in the local FUV field. We additionally study this distribution splitting our sample by spectral type, but we report no change in the trend of \rveil{}-\go{}.

We further investigate the distribution of \rveil{} as a function of binned FUV flux for the NIR excess sources, following the same binning as in Fig. \ref{fig:binned_veil_g0}. We note a median veiling of $\sim 0.3$ for $\log G_{0}<5$, higher than the median \rveil{} for the global population, and only two sources for $5.37<\log G_{0}<5.84$. The number of stars per bin ($\mathrm{N_*}$) for the two highest FUV flux bins is too small ($\mathrm{N_*}\leq4$) to draw more quantitative conclusions. In addition, we examine the distribution of \rveil{} as a function of FUV flux for early-M-type stars and for strongly veiled sources (\rveil{} $> 1$). In both cases, we find no significant evidence that veiling is influenced by the strength of the FUV radiation field. We attribute this lack of correlation primarily to limitations in the adopted veiling modeling. Veiling is inherently wavelength-dependent, and approximating it as a constant across the spectrum can introduce systematic uncertainties in the derived \rveil{} values. In particular, \cinn{} note that this approach may fail to recover heavily veiled sources with \rveil{}>2, and that measurements with \rveil{}>1 are generally less reliable. This reveals that we need a new and more sophisticated way for representing veiling \citep[][\domi{}]{cinn3}

\subsection{Dependence of NIR excess fraction to the local FUV field \label{sec:nir_go}}

\begin{figure}[h!]
    \centering
    \includegraphics[width=0.5\textwidth]{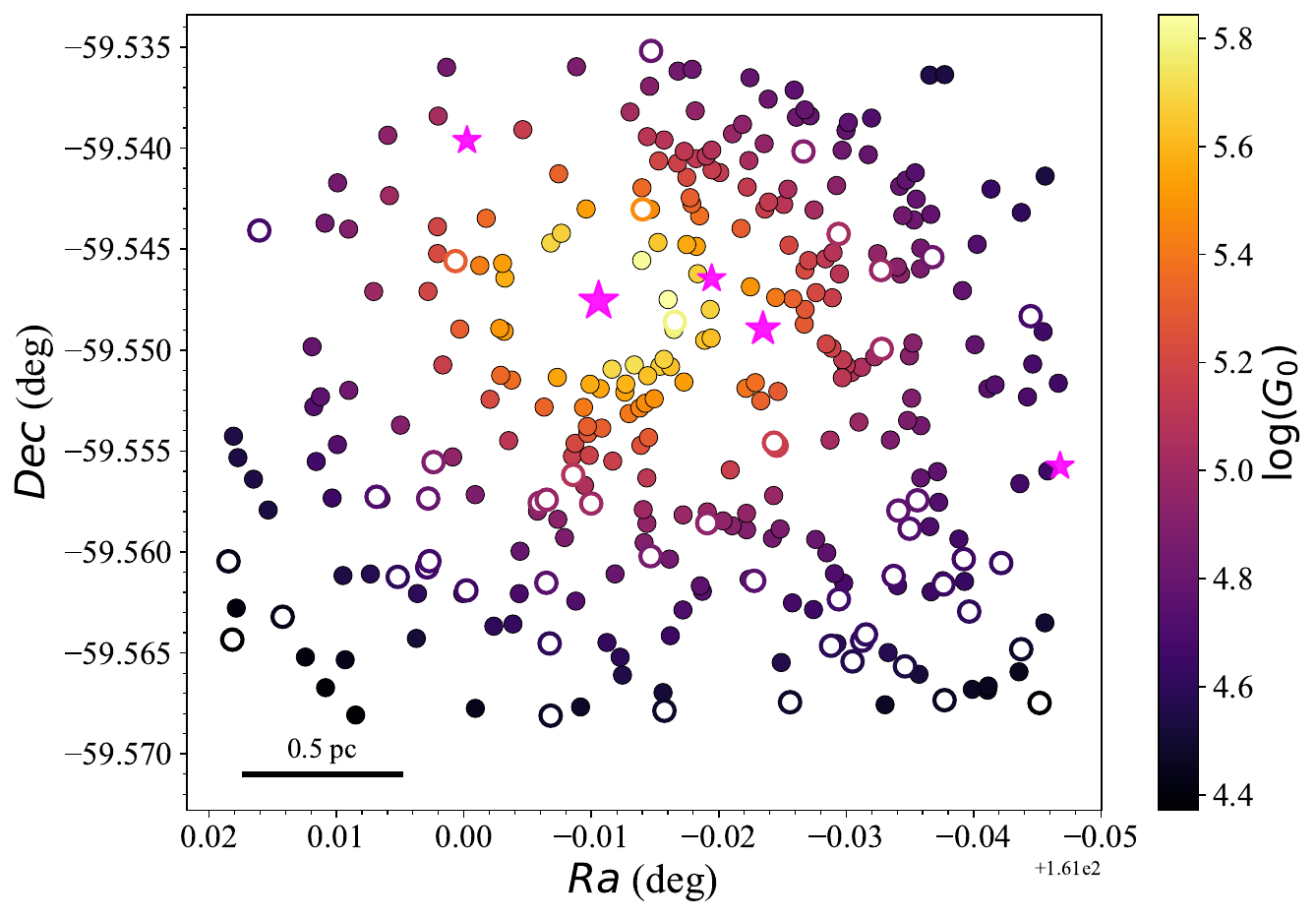}
    \caption{Spatial distribution of the center of Tr14, colored based on the FUV flux strength. With magenta stars we indicate the position of the O-type stars in the region. With empty circles, we highlight the position of stars with 
    NIR excess. \label{fig:spatial+nir}}
    
\end{figure}

In Fig. \ref{fig:spatial+nir}, we display the spatial distribution of our sample, colored according to $\log G_{\mathrm{0}}$. We indicate with empty points the NIR excess sources identified using the \citet{nir} reddening slope, and we include the positions of the O-type stars (taken from \citet{otr14}, with spectral typing from \citet{sota2014}). The five O-type stars seen in Fig.~\ref{fig:spatial+nir} from left to rigt, are: \textit{LS 1823}, an O7V star, \textit{HD 93129}, a visual triple system with one O2 If+O3III interferometric binary and a single O3.5V star, an O8.5V star \citep[Tr14-9 in][]{otr14}, \textit{HD 93128}, an O3.5V spectroscopic binary, and \textit{ALS 15207}, an O9V spectroscopic binary.

From the spatial distribution of the FUV field it is seen that \textit{ALS 15207} (in the far right) doesn't contribute to the expected irradiation of its surrounding low-mass stars. We attribute this effect to the assumed 3D spherical geometry of the region. In addition, we see that the strongest irradiation originates from the central \textit{HD 93129}. The system is one of the earliest and hottest known O-type stars in the Milky Way \citep{,sota2014, gruner2019}, establishing an extreme FUV field.

To better illustrate how the background FUV radiation influences disk evolution in our stellar sample, we examine the fraction of stars exhibiting NIR excess across different bins of $\log G_{\text{0}}$. We define the fraction for each bin separately as:
\begin{equation}
    f = \dfrac{N_{\mathrm{NIR(excess)}}}{N_{*}},\label{equ:frac}
\end{equation}
where $\mathrm{N_{*}}$ is the number of stars in the bin. We derive the uncertainties for $f$ via error propagation using the two Poisson distribution errors:
\begin{equation}
    \delta\mathrm{N_{NIR(excess)}} = \pm \sqrt{\mathrm{N_{NIR(excess)}}} ,\label{equ:poiss_err1}
\end{equation}
\begin{equation}
    \delta\mathrm{N_{*}} = \pm \sqrt{\mathrm{N_{*}}} \label{equ:poiss_err2},
\end{equation}

\begin{figure}[h]
    \centering
    \includegraphics[scale = 0.4]{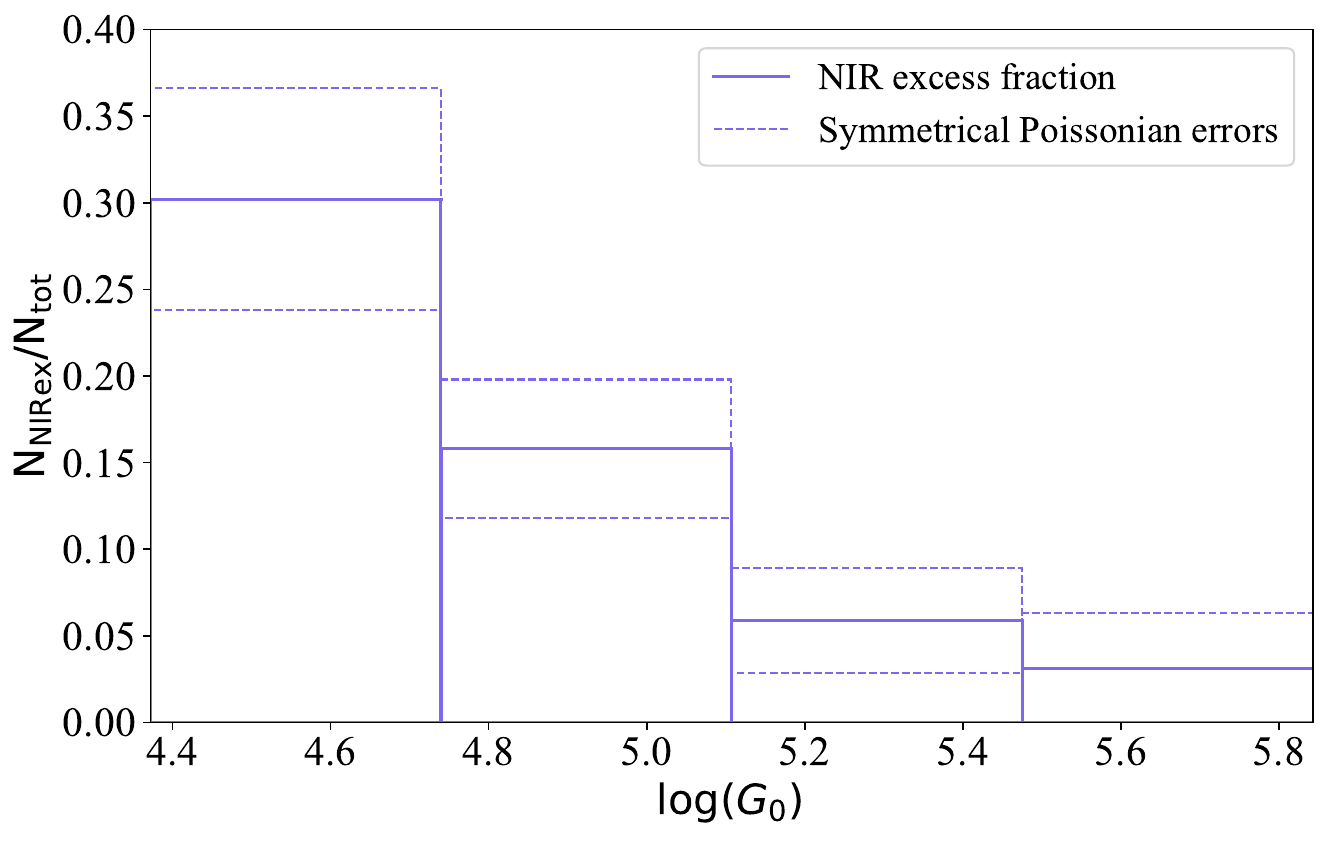}
    \caption{Bar diagram of the NIR excess fraction in 4 different FUV flux bins. The FUV flux is displayed in logarithmic scale and in units of \go{}. The width of each bar indicates the range of each FUV flux bin. The uncertainties derived via error propagation from the Poissonian errors are displayed as dashed lines. \label{fig:nir_frac}}
    
\end{figure}

In Fig. \ref{fig:nir_frac}, we display the NIR excess fraction in the same four FUV flux bins as in Fig.~\ref{fig:binned_veil_g0}. The uncertainties computed are displayed symmetrically for each bin. We find that at an age of $\sim 0.7$ Myr, the disk fraction in the center of Tr14 decreases rapidly from an initial 30\% at low local irradiation, to 3\% for $\log\mathrm{G_0}>5.6$.

\subsection{Tr14 in context with similar regions \label{sec:discuss_disk_frac}}

In the study of \citet{preibis2}, they display their calculated NIR excess fraction for the various clusters in the CNC versus age and accompanied by the literature values for NGC 2024, Taurus, the ONC, IC 348, eta Cha, and Ori OB 1a and 1b. They find that for clusters of the same age, the disk fraction varies. Specifically for the clusters in the CNC, all disk fractions are lower compared to others of the same age, a characteristic attributed to their harsh environment. Nevertheless, for the CNC clusters studied by \citet{preibis1,preibis2}, there is no quantitative analysis of their disk fractions compared to their FUV field.

To place the decreasing disk fraction within the local FUV field of Tr14's center into context, we compare it with the disk fractions in NGC 6611 and Cyg OB2 \citep[see][]{guarc2007, guarc2009, guarc2010, guarc2013, guarcello}. For both regions, the aforementioned works studied in depth the disk fraction in different FUV fluxes. Assuming that regions of the same age and mass distribution will be at the same evolutionary stage, we can expect that the incident FUV field will be the main drive for different disk evolution.

\begin{figure}[h]
    \centering
    \includegraphics[scale = 0.41]{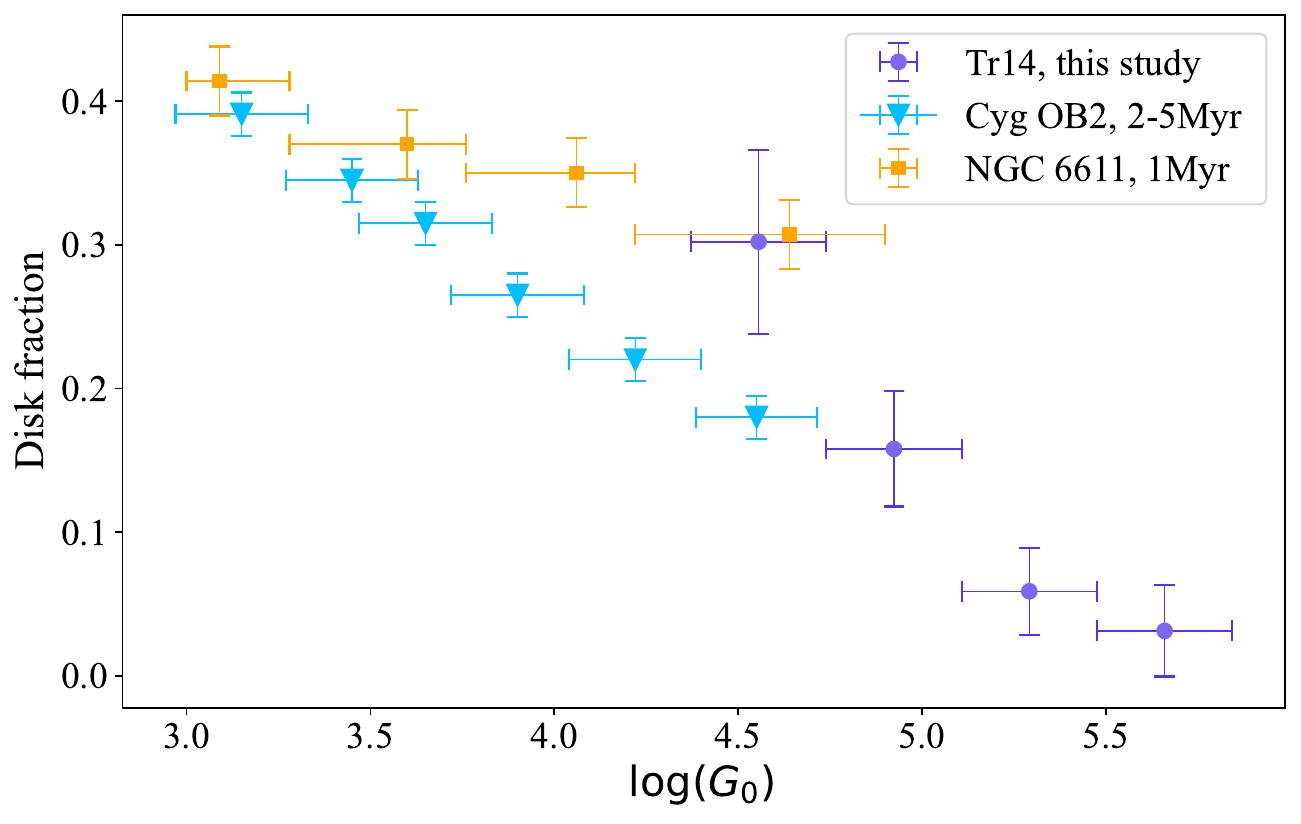}
    \caption{Disk fractions from this study compared with NGC6611 and Cyg OB2. The disk fractions from this study are displayed in 4 FUV flux bins as purple points. The disk fractions from the other clusters are taken from \citet{guarcello}. In all cases, the error bars on the x-axis indicate the range of FUV flux values included in each bin used to calculate the disk fraction. \label{fig:disk_frac+clusters}}
    
\end{figure}

NGC 6611, is an open cluster located in the Eagle Nebula at a distance of 1.75 kpc, and with an estimated age of $\sim1$ Myr. The cluster contains 13 O-type stars, exhibiting an intense UV field. After initially inferring a global disk fraction of 24\% \citep{guarc2009}, varying with the local incident UV flux, \citet[][hereafter \guarc{}]{guarcello}  later recalculated this variation. Based on the 2D separation from the O stars, \guarc{} find the FUV flux in NGC 6611 ranging within $\log \mathrm{G_0} = [3.0, 4.9]$. For $\log \mathrm{G_0}\sim3$, they derive a disk fraction of approximately 41\%, which decreases to 31\% for $\log\mathrm{G_0}\sim4.6$. We present our disk fractions in different FUV flux bins together with the disk fractions measured in NGC 6611 in Fig. \ref{fig:disk_frac+clusters}. The FUV flux is displayed in logarithmic scale and in units of \go{}. We find that our disk fraction in the overlapping range ($\log\mathrm{G_0}=4.4-4.7$) is consistent with the value reported by \guarc{} for NGC 6611, of approximately 30\%. Given the similar estimated ages of Tr14 and NGC 6611, we conclude that the low disk fraction in Tr14 is indeed a result of the intense incident UV flux in the region.

We also compare this result with Cyg OB2. \guarc{} also investigated the association and how external photoevaporation induced by the O-type stars affects the disk fractions. This region is located at half the distance of Tr14 \citep[at approximately 1.4 kpc,][]{rygl} and hosts a large population of low-mass stars. The age of the region has been estimated to be 2-5 Myr old \citep{wright2010, hanson2003, drew2008}, making Cyg OB2 significantly older than Tr14. \guarc{} estimated the FUV flux based on the 2D distances from the OB stars, therefore providing upper limits for the incident radiation. Their sample spans in $\log \mathrm{G_0} = [3.0, 4.7]$, and when divided into bins, they found disk fractions ranging from 40\% at $\log \mathrm{G_0}\sim3$ to 18\% at $\log\mathrm{G_0}\sim4.5$. Since 2D distances underestimate the true distances between stars in three-dimensional space, they simulated a more realistic spatial configuration for Cyg OB2 and recalculated the FUV flux using the 3D representation. They reported little to no difference in the derived disk fractions between the two spatial configurations, considering the uncertainties. We therefore adopt the disk fractions calculated using the 2D distances for comparison with our study. We find that Cyg OB2 exhibits lower disk fractions compared to those found in Tr14 (also displayed in Fig.~\ref{fig:disk_frac+clusters}). In \guarc{}, the authors estimate the average age of the sources within each FUV flux bin. With average ages ranging from 2.3 to 3.6 Myr, Cyg OB2 the lower disk fractions are consistent with the expected decrease of disk survival with age. 

While comparing the three regions, we find clear differences in the dependence of disk fraction on the local FUV radiation field. Even between Tr14 and NGC 6611, which have comparable estimated ages, the decline in disk fraction with increasing FUV flux is considerably steeper in Tr14. In the core of Tr14, the disk fraction decreases by nearly a factor of two between successive FUV flux bins, whereas in NGC 6611 it decreases by only $\sim10\%$ over the range $\log\mathrm{G_0}\approx3\ \text{to}\ 4.5$. This steeper decline may reflect the particularly crowded and intense environment in the core of Tr14. Some models \citep[e.g., ][]{weder2026} indicate that disk dispersal driven by external irradiation does not proceed linearly, implying that a different slope could appear at higher FUV fluxes. Moreover, since comparable measurements are not available for the central regions of NGC 6611 and Cyg OB2, it is not possible to determine whether a similar behavior is present in those clusters. A more direct comparison will require dedicated studies of the cluster cores, both from a theoretical and an observational perspective, to quantify the relation between the strong local irradiation and the disk fraction. For Tr14, extending this analysis to the outer regions of the cluster would allow Fig.~\ref{fig:disk_frac+clusters} to be expanded toward the lower FUV flux regime, providing a more complete view of the dependence of disk fraction on the local radiation field.

\section{Summary and conclusions \label{sec:summ}} 

In this work, we presented a comprehensive spectroscopic analysis of the central region of Tr14, one of the most massive and densely populated clusters in the Carina Nebula Complex. We characterized the low-mass stellar population in the cluster's core, a region previously under-sampled due to extreme background contamination, and we explored the impact of intense external FUV radiation on their circumstellar environments. 

Using the \verb|pySpecMUSE| framework, we accounted for the highly variable H{\sc ii} background emission characteristic of the CNC. We retrieved a sample of 1,615 initial spectra from MUSE/VLT archival observations, which we subsequently refined to a final catalog of 421 spectra with sufficient signal-to-noise for stellar characterization. The catalog represents an increase of 2.8 times in detected sources within the cluster's core compared to previous studies \citepalias{domi}. To derive stellar parameters, we employed the \texttt{SAPSAL-v2} deep learning toolkit \citepalias{cinn2}. Based on a conditional invertible neural network (cINN), this approach enabled the simultaneous determination of the stellar effective temperature (\teff{}), surface gravity (\logg), extinction (\Av), and a simplistic estimate of the optical veiling (\rveil).

Following the cross-match of our sample with near-infrared catalogs \citep{preibis1,preibis2,preibis3}, we computed bolometric luminosities and placed the stars on the HR diagram. Using PARSEC evolutionary tracks \citep{parsec}, we identified 310 stars as bona fide cluster members with ages less than $5 \text{ Myr}$. We estimated an average age of $0.66_{-0.35}^{+0.73}\ \text{Myr}$ for the cluster's core. This is consistent with Tr14 being among the youngest star-forming sites in the region. The average extinction in the central region is $A_V = 2.43 \pm 0.67 \text{ mag}$, consistent with literature values for the CNC.

Moreover, we studied and quantified the local external FUV radiation field in units of \go{} using a 3D statistical approach \citep[based on ][]{g0}. We found that the low-mass population is exposed to intense radiation, with values ranging from $10^{4.2}$\go{} to $10^{6.0}$\go{}, considering the uncertainties. These values indicate that the core of Tr14 provides an environment where external photoevaporation may significantly influence circumstellar disk evolution.

To quantify the effect of external photoevaporation, we first investigated the relation between optical veiling and the local FUV radiation field. We found no significant correlation between $r_{\rm veil}$ and $G_0$. Given the limitations of assuming a wavelength-independent veiling component and the uncertainties at high veiling values, we could not draw firm conclusions on the impact of external photoevaporation on accretion-related observables from the current dataset.

Next, we studied the impact of the FUV radiation by comparing the spatial distribution of stars with inner disks, to the local FUV flux. Using the NIR photometry from the HAWK-I and VISTA catalogs \citep{preibis1, preibis2, preibis3}, we derive 17\% of our stellar sample as NIR excess sources, with the majority of disk-bearing sources located farther from O-type stars in the region. Specifically, we found that at low local irradiation ($\log G_{\text{0}}\sim 4.6$), 30\% of stars exhibit NIR excess, suggesting the presence of circumstellar disks, while this fraction decreases to 5\% for $\log G_{\text{0}}>5$. 

To put our region into perspective, we compared the disk fractions in Tr14 with those measured in NGC 6611 and Cyg OB2 by \guarc{}. We found that the disk fraction in Tr14 is consistent with that of NGC 6611, a young cluster with estimated age of $\sim1$ Myr. At $\log G_{\text{0}}\sim 4.6$, both clusters showed a disk fraction of approximately 30\%, while the older Cyg OB2 displays a lower disk fraction of 18\%. In this context, our results are consistent with the expected dependence of disk survival on both environmental irradiation and cluster age.

Overall, our results suggested that the strong FUV radiation field in the core of Tr14 causes a reduced disk fraction, consistent with the expected role of external photoevaporation in shaping disk evolution in massive star-forming environments. The combination of IFU spectroscopy, dedicated extraction methods, and machine-learning-based stellar characterization provided a pathway for investigating stellar populations and disk evolution in distant massive star-forming regions.

\section*{Data availability}
The stellar-population-analysis products generated for this study are available from the corresponding author upon request.

\begin{acknowledgements}

This work was partly supported by the Italian Ministero dell Istruzione, Università e Ricerca through the grant Progetti Premiali 2012 -– iALMA (CUP C52I13000140001). This project has received funding from the European Research Council (ERC) via the ERC Synergy Grant ECOGAL (grant 855130). Views and opinions expressed are however those of the author(s) only and do not necessarily reflect those of the European Union or the European Research Council Executive Agency. Neither the European Union nor the granting authority can be held responsible for them. 

DI acknowledges support from collaborations and/or information ex- change within NASA’s Nexus for Exoplanet System Science (NExSS) research coordination network sponsored by NASA’s Science Mission Directorate under Agreement No. 80NSSC21K0593 for the program “Alien Earths”.

RSK acknowledges financial support from the ERC via Synergy Grant ``ECOGAL'' (project ID 855130) and from the German Excellence Strategy via the Heidelberg Cluster ``STRUCTURES'' (EXC 2181 - 390900948). In addition RSK is grateful for funding from the German Ministry for Economy and Energy (BMWE) in project ``MAINN'' (funding ID 50OO2206), and from DFG and ANR for project ``STARCLUSTERS'' (funding ID KL 1358/22-1).

TP acknowledges support from the Excellence Cluster ORIGINS which is funded by the Deutsche Forschungsgemeinschaft (DFG, German Research Foundation) under Germany's Excellence Strategy - EXC-2094 - 390783311.
\end{acknowledgements}

\bibliographystyle{aa}
\bibliography{bib}

@ARTICLE{domi,
       author = {{Itrich}, Dominika and {Testi}, Leonardo and {Beccari}, Giacomo and {Manara}, Carlo F. and {Reiter}, Megan and {Preibisch}, Thomas and {McLeod}, Anna F. and {Rosotti}, Giovanni and {Klessen}, Ralf and {Molinari}, Sergio and {Hennebelle}, Patrick},
        title = "{The population of young low-mass stars in Trumpler 14}",
      journal = {\aap},
         year = 2024,
        month = may,
       volume = {685},
          eid = {A100},
        pages = {A100},
          doi = {10.1051/0004-6361/202347380},
archivePrefix = {arXiv},
       eprint = {2309.14168},
 primaryClass = {astro-ph.SR},
       adsurl = {https://ui.adsabs.harvard.edu/abs/2024A&A...685A.100I}
}

@ARTICLE{wfi,
       author = {{Beccari}, G. and {De Marchi}, G. and {Panagia}, N. and {Valenti}, E. and {Carraro}, G. and {Romaniello}, M. and {Zoccali}, M. and {Weidner}, C.},
        title = "{Mass accretion rates from multiband photometry in the Carina Nebula: the case of Trumpler 14}",
      journal = {\aap},
         year = 2015,
        month = jan,
       volume = {574},
          eid = {A44},
        pages = {A44},
          doi = {10.1051/0004-6361/201424077},
archivePrefix = {arXiv},
       eprint = {1409.4370},
 primaryClass = {astro-ph.SR},
       adsurl = {https://ui.adsabs.harvard.edu/abs/2015A&A...574A..44B}
}

@ARTICLE{reduce,
       author = {{Weilbacher}, Peter M. and {Palsa}, Ralf and {Streicher}, Ole and {Bacon}, Roland and {Urrutia}, Tanya and {Wisotzki}, Lutz and {Conseil}, Simon and {Husemann}, Bernd and {Jarno}, Aur{\'e}lien and {Kelz}, Andreas and {P{\'e}contal-Rousset}, Arlette and {Richard}, Johan and {Roth}, Martin M. and {Selman}, Fernando and {Vernet}, Jo{\"e}l},
        title = "{The data processing pipeline for the MUSE instrument}",
      journal = {\aap},
         year = 2020,
        month = sep,
       volume = {641},
          eid = {A28},
        pages = {A28},
          doi = {10.1051/0004-6361/202037855},
archivePrefix = {arXiv},
       eprint = {2006.08638},
 primaryClass = {astro-ph.IM},
       adsurl = {https://ui.adsabs.harvard.edu/abs/2020A&A...641A..28W}
}

@ARTICLE{esorex,
       author = {{Freudling}, W. and {Romaniello}, M. and {Bramich}, D.~M. and {Ballester}, P. and {Forchi}, V. and {Garc{\'\i}a-Dabl{\'o}}, C.~E. and {Moehler}, S. and {Neeser}, M.~J.},
        title = "{Automated data reduction workflows for astronomy. The ESO Reflex environment}",
      journal = {\aap},
         year = 2013,
        month = nov,
       volume = {559},
          eid = {A96},
        pages = {A96},
          doi = {10.1051/0004-6361/201322494},
archivePrefix = {arXiv},
       eprint = {1311.5411},
 primaryClass = {astro-ph.IM},
       adsurl = {https://ui.adsabs.harvard.edu/abs/2013A&A...559A..96F}
}

@ARTICLE{gaia1,
       author = {{Gaia Collaboration} and {Prusti}, T. and {de Bruijne}, J.~H.~J. and {Brown}, A.~G.~A. and {Vallenari}, A. and {Babusiaux}, C. and {Bailer-Jones}, C.~A.~L. and {Bastian}, U. and {Biermann}, M. and {Evans}, D.~W. and {Eyer}, L. and {Jansen}, F. and {Jordi}, C. and {Klioner}, S.~A. and {Lammers}, U. and {Lindegren}, L. and {Luri}, X. and {Mignard}, F. and {Milligan}, D.~J. and {Panem}, C. and {Poinsignon}, V. and {Pourbaix}, D. and {Randich}, S. and {Sarri}, G. and {Sartoretti}, P. and {Siddiqui}, H.~I. and {Soubiran}, C. and {Valette}, V. and {van Leeuwen}, F. and {Walton}, N.~A. and {Aerts}, C. and {Arenou}, F. and {Cropper}, M. and {Drimmel}, R. and {H{\o}g}, E. and {Katz}, D. and {Lattanzi}, M.~G. and {O'Mullane}, W. and {Grebel}, E.~K. and {Holland}, A.~D. and {Huc}, C. and {Passot}, X. and {Bramante}, L. and {Cacciari}, C. and {Casta{\~n}eda}, J. and {Chaoul}, L. and {Cheek}, N. and {De Angeli}, F. and {Fabricius}, C. and {Guerra}, R. and {Hern{\'a}ndez}, J. and {Jean-Antoine-Piccolo}, A. and {Masana}, E. and {Messineo}, R. and {Mowlavi}, N. and {Nienartowicz}, K. and {Ord{\'o}{\~n}ez-Blanco}, D. and {Panuzzo}, P. and {Portell}, J. and {Richards}, P.~J. and {Riello}, M. and {Seabroke}, G.~M. and {Tanga}, P. and {Th{\'e}venin}, F. and {Torra}, J. and {Els}, S.~G. and {Gracia-Abril}, G. and {Comoretto}, G. and {Garcia-Reinaldos}, M. and {Lock}, T. and {Mercier}, E. and {Altmann}, M. and {Andrae}, R. and {Astraatmadja}, T.~L. and {Bellas-Velidis}, I. and {Benson}, K. and {Berthier}, J. and {Blomme}, R. and {Busso}, G. and {Carry}, B. and {Cellino}, A. and {Clementini}, G. and {Cowell}, S. and {Creevey}, O. and {Cuypers}, J. and {Davidson}, M. and {De Ridder}, J. and {de Torres}, A. and {Delchambre}, L. and {Dell'Oro}, A. and {Ducourant}, C. and {Fr{\'e}mat}, Y. and {Garc{\'\i}a-Torres}, M. and {Gosset}, E. and {Halbwachs}, J. -L. and {Hambly}, N.~C. and {Harrison}, D.~L. and {Hauser}, M. and {Hestroffer}, D. and {Hodgkin}, S.~T. and {Huckle}, H.~E. and {Hutton}, A. and {Jasniewicz}, G. and {Jordan}, S. and {Kontizas}, M. and {Korn}, A.~J. and {Lanzafame}, A.~C. and {Manteiga}, M. and {Moitinho}, A. and {Muinonen}, K. and {Osinde}, J. and {Pancino}, E. and {Pauwels}, T. and {Petit}, J. -M. and {Recio-Blanco}, A. and {Robin}, A.~C. and {Sarro}, L.~M. and {Siopis}, C. and {Smith}, M. and {Smith}, K.~W. and {Sozzetti}, A. and {Thuillot}, W. and {van Reeven}, W. and {Viala}, Y. and {Abbas}, U. and {Abreu Aramburu}, A. and {Accart}, S. and {Aguado}, J.~J. and {Allan}, P.~M. and {Allasia}, W. and {Altavilla}, G. and {{\'A}lvarez}, M.~A. and {Alves}, J. and {Anderson}, R.~I. and {Andrei}, A.~H. and {Anglada Varela}, E. and {Antiche}, E. and {Antoja}, T. and {Ant{\'o}n}, S. and {Arcay}, B. and {Atzei}, A. and {Ayache}, L. and {Bach}, N. and {Baker}, S.~G. and {Balaguer-N{\'u}{\~n}ez}, L. and {Barache}, C. and {Barata}, C. and {Barbier}, A. and {Barblan}, F. and {Baroni}, M. and {Barrado y Navascu{\'e}s}, D. and {Barros}, M. and {Barstow}, M.~A. and {Becciani}, U. and {Bellazzini}, M. and {Bellei}, G. and {Bello Garc{\'\i}a}, A. and {Belokurov}, V. and {Bendjoya}, P. and {Berihuete}, A. and {Bianchi}, L. and {Bienaym{\'e}}, O. and {Billebaud}, F. and {Blagorodnova}, N. and {Blanco-Cuaresma}, S. and {Boch}, T. and {Bombrun}, A. and {Borrachero}, R. and {Bouquillon}, S. and {Bourda}, G. and {Bouy}, H. and {Bragaglia}, A. and {Breddels}, M.~A. and {Brouillet}, N. and {Br{\"u}semeister}, T. and {Bucciarelli}, B. and {Budnik}, F. and {Burgess}, P. and {Burgon}, R. and {Burlacu}, A. and {Busonero}, D. and {Buzzi}, R. and {Caffau}, E. and {Cambras}, J. and {Campbell}, H. and {Cancelliere}, R. and {Cantat-Gaudin}, T. and {Carlucci}, T. and {Carrasco}, J.~M. and {Castellani}, M. and {Charlot}, P. and {Charnas}, J. and {Charvet}, P. and {Chassat}, F. and {Chiavassa}, A. and {Clotet}, M. and {Cocozza}, G. and {Collins}, R.~S. and {Collins}, P. and {Costigan}, G.},
        title = "{The Gaia mission}",
      journal = {\aap},
         year = 2016,
        month = nov,
       volume = {595},
          eid = {A1},
        pages = {A1},
          doi = {10.1051/0004-6361/201629272},
archivePrefix = {arXiv},
       eprint = {1609.04153},
 primaryClass = {astro-ph.IM},
       adsurl = {https://ui.adsabs.harvard.edu/abs/2016A&A...595A...1G}
}

@ARTICLE{gaia2,
       author = {{Gaia Collaboration} and {Vallenari}, A. and {Brown}, A.~G.~A. and {Prusti}, T. and {de Bruijne}, J.~H.~J. and {Arenou}, F. and {Babusiaux}, C. and {Biermann}, M. and {Creevey}, O.~L. and {Ducourant}, C. and {Evans}, D.~W. and {Eyer}, L. and {Guerra}, R. and {Hutton}, A. and {Jordi}, C. and {Klioner}, S.~A. and {Lammers}, U.~L. and {Lindegren}, L. and {Luri}, X. and {Mignard}, F. and {Panem}, C. and {Pourbaix}, D. and {Randich}, S. and {Sartoretti}, P. and {Soubiran}, C. and {Tanga}, P. and {Walton}, N.~A. and {Bailer-Jones}, C.~A.~L. and {Bastian}, U. and {Drimmel}, R. and {Jansen}, F. and {Katz}, D. and {Lattanzi}, M.~G. and {van Leeuwen}, F. and {Bakker}, J. and {Cacciari}, C. and {Casta{\~n}eda}, J. and {De Angeli}, F. and {Fabricius}, C. and {Fouesneau}, M. and {Fr{\'e}mat}, Y. and {Galluccio}, L. and {Guerrier}, A. and {Heiter}, U. and {Masana}, E. and {Messineo}, R. and {Mowlavi}, N. and {Nicolas}, C. and {Nienartowicz}, K. and {Pailler}, F. and {Panuzzo}, P. and {Riclet}, F. and {Roux}, W. and {Seabroke}, G.~M. and {Sordo}, R. and {Th{\'e}venin}, F. and {Gracia-Abril}, G. and {Portell}, J. and {Teyssier}, D. and {Altmann}, M. and {Andrae}, R. and {Audard}, M. and {Bellas-Velidis}, I. and {Benson}, K. and {Berthier}, J. and {Blomme}, R. and {Burgess}, P.~W. and {Busonero}, D. and {Busso}, G. and {C{\'a}novas}, H. and {Carry}, B. and {Cellino}, A. and {Cheek}, N. and {Clementini}, G. and {Damerdji}, Y. and {Davidson}, M. and {de Teodoro}, P. and {Nu{\~n}ez Campos}, M. and {Delchambre}, L. and {Dell'Oro}, A. and {Esquej}, P. and {Fern{\'a}ndez-Hern{\'a}ndez}, J. and {Fraile}, E. and {Garabato}, D. and {Garc{\'\i}a-Lario}, P. and {Gosset}, E. and {Haigron}, R. and {Halbwachs}, J. -L. and {Hambly}, N.~C. and {Harrison}, D.~L. and {Hern{\'a}ndez}, J. and {Hestroffer}, D. and {Hodgkin}, S.~T. and {Holl}, B. and {Jan{\ss}en}, K. and {Jevardat de Fombelle}, G. and {Jordan}, S. and {Krone-Martins}, A. and {Lanzafame}, A.~C. and {L{\"o}ffler}, W. and {Marchal}, O. and {Marrese}, P.~M. and {Moitinho}, A. and {Muinonen}, K. and {Osborne}, P. and {Pancino}, E. and {Pauwels}, T. and {Recio-Blanco}, A. and {Reyl{\'e}}, C. and {Riello}, M. and {Rimoldini}, L. and {Roegiers}, T. and {Rybizki}, J. and {Sarro}, L.~M. and {Siopis}, C. and {Smith}, M. and {Sozzetti}, A. and {Utrilla}, E. and {van Leeuwen}, M. and {Abbas}, U. and {{\'A}brah{\'a}m}, P. and {Abreu Aramburu}, A. and {Aerts}, C. and {Aguado}, J.~J. and {Ajaj}, M. and {Aldea-Montero}, F. and {Altavilla}, G. and {{\'A}lvarez}, M.~A. and {Alves}, J. and {Anders}, F. and {Anderson}, R.~I. and {Anglada Varela}, E. and {Antoja}, T. and {Baines}, D. and {Baker}, S.~G. and {Balaguer-N{\'u}{\~n}ez}, L. and {Balbinot}, E. and {Balog}, Z. and {Barache}, C. and {Barbato}, D. and {Barros}, M. and {Barstow}, M.~A. and {Bartolom{\'e}}, S. and {Bassilana}, J. -L. and {Bauchet}, N. and {Becciani}, U. and {Bellazzini}, M. and {Berihuete}, A. and {Bernet}, M. and {Bertone}, S. and {Bianchi}, L. and {Binnenfeld}, A. and {Blanco-Cuaresma}, S. and {Blazere}, A. and {Boch}, T. and {Bombrun}, A. and {Bossini}, D. and {Bouquillon}, S. and {Bragaglia}, A. and {Bramante}, L. and {Breedt}, E. and {Bressan}, A. and {Brouillet}, N. and {Brugaletta}, E. and {Bucciarelli}, B. and {Burlacu}, A. and {Butkevich}, A.~G. and {Buzzi}, R. and {Caffau}, E. and {Cancelliere}, R. and {Cantat-Gaudin}, T. and {Carballo}, R. and {Carlucci}, T. and {Carnerero}, M.~I. and {Carrasco}, J.~M. and {Casamiquela}, L. and {Castellani}, M. and {Castro-Ginard}, A. and {Chaoul}, L. and {Charlot}, P. and {Chemin}, L. and {Chiaramida}, V. and {Chiavassa}, A. and {Chornay}, N. and {Comoretto}, G. and {Contursi}, G. and {Cooper}, W.~J. and {Cornez}, T. and {Cowell}, S. and {Crifo}, F. and {Cropper}, M. and {Crosta}, M. and {Crowley}, C. and {Dafonte}, C. and {Dapergolas}, A. and {David}, M. and {David}, P. and {de Laverny}, P. and {De Luise}, F. and {De March}, R.},
        title = "{Gaia Data Release 3. Summary of the content and survey properties}",
      journal = {\aap},
         year = 2023,
        month = jun,
       volume = {674},
          eid = {A1},
        pages = {A1},
          doi = {10.1051/0004-6361/202243940},
archivePrefix = {arXiv},
       eprint = {2208.00211},
 primaryClass = {astro-ph.GA},
       adsurl = {https://ui.adsabs.harvard.edu/abs/2023A&A...674A...1G}
}

@software{photutils,
       author = {{Bradley}, Larry and {Sip{\H{o}}cz}, Brigitta and {Robitaille}, Thomas and {Tollerud}, Erik and {Vin{\'\i}cius}, Z{\'e} and {Deil}, Christoph and {Barbary}, Kyle and {Wilson}, Tom J and {Busko}, Ivo and {Donath}, Axel and {G{\"u}nther}, Hans Moritz and {Cara}, Mihai and {Lim}, P.~L. and {Me{\ss}linger}, Sebastian and {Conseil}, Simon and {Burnett}, Zach and {Bostroem}, Azalee and {Droettboom}, Michael and {Bray}, E.~M. and {Andersen Bratholm}, Lars and {Ginsburg}, Adam and {Jamieson}, William and {Barentsen}, Geert and {Craig}, Matt and {Morris}, Brett M. and {Perrin}, Marshall and {Rathi}, Shivangee and {Pascual}, Sergio and {Georgiev}, Iskren Y.},
        title = "{astropy/photutils: 2.0.2}",
         year = 2024,
        month = oct,
          eid = {10.5281/zenodo.13989456},
          doi = {10.5281/zenodo.13989456},
      version = {2.0.2},
    publisher = {Zenodo},
       adsurl = {https://ui.adsabs.harvard.edu/abs/2024zndo..13989456B}
}

@ARTICLE{daofind,
       author = {{Stetson}, Peter B.},
        title = "{DAOPHOT: A Computer Program for Crowded-Field Stellar Photometry}",
      journal = {\pasp},
         year = 1987,
        month = mar,
       volume = {99},
        pages = {191},
          doi = {10.1086/131977},
       adsurl = {https://ui.adsabs.harvard.edu/abs/1987PASP...99..191S}
}

@ARTICLE{cinn,
       author = {{Kang}, Da Eun and {Ksoll}, Victor F. and {Itrich}, Dominika and {Testi}, Leonardo and {Klessen}, Ralf S. and {Hennebelle}, Patrick and {Molinari}, Sergio},
        title = "{Spectral classification of young stars using conditional invertible neural networks. I. Introducing and validating the method}",
      journal = {\aap},
         year = 2023,
        month = jun,
       volume = {674},
          eid = {A175},
        pages = {A175},
          doi = {10.1051/0004-6361/202346345},
archivePrefix = {arXiv},
       eprint = {2304.08398},
 primaryClass = {astro-ph.SR},
       adsurl = {https://ui.adsabs.harvard.edu/abs/2023A&A...674A.175K}
}

@ARTICLE{cinn2,
       author = {{Kang}, Da Eun and {Itrich}, Dominika and {Ksoll}, Victor F. and {Testi}, Leonardo and {Klessen}, Ralf S. and {Molinari}, Sergio},
        title = "{Spectral classification of young stars using conditional invertible neural networks: II. Application to Trumpler 14 in Carina}",
      journal = {\aap},
         year = 2025,
        month = may,
       volume = {697},
          eid = {A39},
        pages = {A39},
          doi = {10.1051/0004-6361/202450394},
archivePrefix = {arXiv},
       eprint = {2503.19697},
 primaryClass = {astro-ph.SR},
       adsurl = {https://ui.adsabs.harvard.edu/abs/2025A&A...697A..39K}
}

@INPROCEEDINGS{allard1,
       author = {{Allard}, F. and {Homeier}, D. and {Freytag}, B.},
        title = "{Model Atmospheres From Very Low Mass Stars to Brown Dwarfs}",
    booktitle = {16th Cambridge Workshop on Cool Stars, Stellar Systems, and the Sun},
         year = 2011,
       editor = {{Johns-Krull}, Christopher and {Browning}, Matthew K. and {West}, Andrew A.},
       series = {Astronomical Society of the Pacific Conference Series},
       volume = {448},
        month = dec,
        pages = {91},
          doi = {10.48550/arXiv.1011.5405},
archivePrefix = {arXiv},
       eprint = {1011.5405},
 primaryClass = {astro-ph.SR},
       adsurl = {https://ui.adsabs.harvard.edu/abs/2011ASPC..448...91A}
}

@INPROCEEDINGS{allard2,
       author = {{Allard}, F. and {Homeier}, D. and {Freytag}, B. and {Sharp}, C.~M.},
        title = "{Atmospheres From Very Low-Mass Stars to Extrasolar Planets}",
    booktitle = {EAS Publications Series},
         year = 2012,
       editor = {{Reyl{\'e}}, C{\'e}line and {Charbonnel}, Corinne and {Schultheis}, Mathias},
       series = {EAS Publications Series},
       volume = {57},
        month = nov,
    publisher = {EDP},
        pages = {3-43},
          doi = {10.1051/eas/1257001},
archivePrefix = {arXiv},
       eprint = {1206.1021},
 primaryClass = {astro-ph.SR},
       adsurl = {https://ui.adsabs.harvard.edu/abs/2012EAS....57....3A}
}

@ARTICLE{cardeli,
       author = {{Cardelli}, Jason A. and {Clayton}, Geoffrey C. and {Mathis}, John S.},
        title = "{The Relationship between Infrared, Optical, and Ultraviolet Extinction}",
      journal = {\apj},
         year = 1989,
        month = oct,
       volume = {345},
        pages = {245},
          doi = {10.1086/167900},
       adsurl = {https://ui.adsabs.harvard.edu/abs/1989ApJ...345..245C}
}

@ARTICLE{hur,
       author = {{Hur}, Hyeonoh and {Sung}, Hwankyung and {Bessell}, Michael S.},
        title = "{Distance and the Initial Mass Function of Young Open Clusters in the {\ensuremath{\eta}} Carina Nebula: Tr 14 and Tr 16}",
      journal = {\aj},
         year = 2012,
        month = feb,
       volume = {143},
       number = {2},
          eid = {41},
        pages = {41},
          doi = {10.1088/0004-6256/143/2/41},
archivePrefix = {arXiv},
       eprint = {1201.0623},
 primaryClass = {astro-ph.SR},
       adsurl = {https://ui.adsabs.harvard.edu/abs/2012AJ....143...41H}
}

@ARTICLE{preibis1,
       author = {{Preibisch}, Thomas and {Hodgkin}, Simon and {Irwin}, Mike and {Lewis}, James R. and {King}, Robert R. and {McCaughrean}, Mark J. and {Zinnecker}, Hans and {Townsley}, Leisa and {Broos}, Patrick},
        title = "{Near-infrared Properties of the X-ray-emitting Young Stellar Objects in the Carina Nebula}",
      journal = {\apjs},
         year = 2011,
        month = may,
       volume = {194},
       number = {1},
          eid = {10},
        pages = {10},
          doi = {10.1088/0067-0049/194/1/10},
archivePrefix = {arXiv},
       eprint = {1103.2052},
 primaryClass = {astro-ph.SR},
       adsurl = {https://ui.adsabs.harvard.edu/abs/2011ApJS..194...10P}
}

@ARTICLE{preibis2,
       author = {{Preibisch}, T. and {Ratzka}, T. and {Kuderna}, B. and {Ohlendorf}, H. and {King}, R.~R. and {Hodgkin}, S. and {Irwin}, M. and {Lewis}, J.~R. and {McCaughrean}, M.~J. and {Zinnecker}, H.},
        title = "{Deep wide-field near-infrared survey of the Carina Nebula}",
      journal = {\aap},
         year = 2011,
        month = jun,
       volume = {530},
          eid = {A34},
        pages = {A34},
          doi = {10.1051/0004-6361/201116781},
archivePrefix = {arXiv},
       eprint = {1104.3477},
 primaryClass = {astro-ph.GA},
       adsurl = {https://ui.adsabs.harvard.edu/abs/2011A&A...530A..34P}
}

@ARTICLE{preibis3,
       author = {{Preibisch}, T. and {Zeidler}, P. and {Ratzka}, T. and {Roccatagliata}, V. and {Petr-Gotzens}, M.~G.},
        title = "{The VISTA Carina Nebula Survey . I. Introduction and source catalog}",
      journal = {\aap},
         year = 2014,
        month = dec,
       volume = {572},
          eid = {A116},
        pages = {A116},
          doi = {10.1051/0004-6361/201424045},
       adsurl = {https://ui.adsabs.harvard.edu/abs/2014A&A...572A.116P}
}

@ARTICLE{kh,
       author = {{Kenyon}, Scott J. and {Hartmann}, Lee},
        title = "{Pre-Main-Sequence Evolution in the Taurus-Auriga Molecular Cloud}",
      journal = {\apjs},
         year = 1995,
        month = nov,
       volume = {101},
        pages = {117},
          doi = {10.1086/192235},
       adsurl = {https://ui.adsabs.harvard.edu/abs/1995ApJS..101..117K}
}

@ARTICLE{lhu,
       author = {{Luhman}, K.~L. and {Stauffer}, John R. and {Muench}, A.~A. and {Rieke}, G.~H. and {Lada}, E.~A. and {Bouvier}, J. and {Lada}, C.~J.},
        title = "{A Census of the Young Cluster IC 348}",
      journal = {\apj},
         year = 2003,
        month = aug,
       volume = {593},
       number = {2},
        pages = {1093-1115},
          doi = {10.1086/376594},
archivePrefix = {arXiv},
       eprint = {astro-ph/0304409},
 primaryClass = {astro-ph},
       adsurl = {https://ui.adsabs.harvard.edu/abs/2003ApJ...593.1093L}
}

@ARTICLE{parsec,
       author = {{Bressan}, Alessandro and {Marigo}, Paola and {Girardi}, L{\'e}o. and {Salasnich}, Bernardo and {Dal Cero}, Claudia and {Rubele}, Stefano and {Nanni}, Ambra},
        title = "{PARSEC: stellar tracks and isochrones with the PAdova and TRieste Stellar Evolution Code}",
      journal = {\mnras},
         year = 2012,
        month = nov,
       volume = {427},
       number = {1},
        pages = {127-145},
          doi = {10.1111/j.1365-2966.2012.21948.x},
archivePrefix = {arXiv},
       eprint = {1208.4498},
 primaryClass = {astro-ph.SR},
       adsurl = {https://ui.adsabs.harvard.edu/abs/2012MNRAS.427..127B}
}

@ARTICLE{miler,
       author = {{Miller}, G.~E. and {Scalo}, J.~M.},
        title = "{On the birthplaces of stars.}",
      journal = {\pasp},
         year = 1978,
        month = oct,
       volume = {90},
        pages = {506-513},
          doi = {10.1086/130373},
       adsurl = {https://ui.adsabs.harvard.edu/abs/1978PASP...90..506M}
}

@ARTICLE{astropy2022,
       author = {{Astropy Collaboration} and {Price-Whelan}, Adrian M. and {Lim}, Pey Lian and {Earl}, Nicholas and {Starkman}, Nathaniel and {Bradley}, Larry and {Shupe}, David L. and {Patil}, Aarya A. and {Corrales}, Lia and {Brasseur}, C.~E. and {N{"o}the}, Maximilian and {Donath}, Axel and {Tollerud}, Erik and {Morris}, Brett M. and {Ginsburg}, Adam and {Vaher}, Eero and {Weaver}, Benjamin A. and {Tocknell}, James and {Jamieson}, William and {van Kerkwijk}, Marten H. and {Robitaille}, Thomas P. and {Merry}, Bruce and {Bachetti}, Matteo and {G{"u}nther}, H. Moritz and {Aldcroft}, Thomas L. and {Alvarado-Montes}, Jaime A. and {Archibald}, Anne M. and {B{'o}di}, Attila and {Bapat}, Shreyas and {Barentsen}, Geert and {Baz{'a}n}, Juanjo and {Biswas}, Manish and {Boquien}, M{'e}d{'e}ric and {Burke}, D.~J. and {Cara}, Daria and {Cara}, Mihai and {Conroy}, Kyle E. and {Conseil}, Simon and {Craig}, Matthew W. and {Cross}, Robert M. and {Cruz}, Kelle L. and {D'Eugenio}, Francesco and {Dencheva}, Nadia and {Devillepoix}, Hadrien A.~R. and {Dietrich}, J{"o}rg P. and {Eigenbrot}, Arthur Davis and {Erben}, Thomas and {Ferreira}, Leonardo and {Foreman-Mackey}, Daniel and {Fox}, Ryan and {Freij}, Nabil and {Garg}, Suyog and {Geda}, Robel and {Glattly}, Lauren and {Gondhalekar}, Yash and {Gordon}, Karl D. and {Grant}, David and {Greenfield}, Perry and {Groener}, Austen M. and {Guest}, Steve and {Gurovich}, Sebastian and {Handberg}, Rasmus and {Hart}, Akeem and {Hatfield-Dodds}, Zac and {Homeier}, Derek and {Hosseinzadeh}, Griffin and {Jenness}, Tim and {Jones}, Craig K. and {Joseph}, Prajwel and {Kalmbach}, J. Bryce and {Karamehmetoglu}, Emir and {Ka{l}uszy{'n}ski}, Miko{l}aj and {Kelley}, Michael S.~P. and {Kern}, Nicholas and {Kerzendorf}, Wolfgang E. and {Koch}, Eric W. and {Kulumani}, Shankar and {Lee}, Antony and {Ly}, Chun and {Ma}, Zhiyuan and {MacBride}, Conor and {Maljaars}, Jakob M. and {Muna}, Demitri and {Murphy}, N.~A. and {Norman}, Henrik and {O'Steen}, Richard and {Oman}, Kyle A. and {Pacifici}, Camilla and {Pascual}, Sergio and {Pascual-Granado}, J. and {Patil}, Rohit R. and {Perren}, Gabriel I. and {Pickering}, Timothy E. and {Rastogi}, Tanuj and {Roulston}, Benjamin R. and {Ryan}, Daniel F. and {Rykoff}, Eli S. and {Sabater}, Jose and {Sakurikar}, Parikshit and {Salgado}, Jes{'u}s and {Sanghi}, Aniket and {Saunders}, Nicholas and {Savchenko}, Volodymyr and {Schwardt}, Ludwig and {Seifert-Eckert}, Michael and {Shih}, Albert Y. and {Jain}, Anany Shrey and {Shukla}, Gyanendra and {Sick}, Jonathan and {Simpson}, Chris and {Singanamalla}, Sudheesh and {Singer}, Leo P. and {Singhal}, Jaladh and {Sinha}, Manodeep and {Sip{H{o}}cz}, Brigitta M. and {Spitler}, Lee R. and {Stansby}, David and {Streicher}, Ole and {{{S}}umak}, Jani and {Swinbank}, John D. and {Taranu}, Dan S. and {Tewary}, Nikita and {Tremblay}, Grant R. and {Val-Borro}, Miguel de and {Van Kooten}, Samuel J. and {Vasovi{'c}}, Zlatan and {Verma}, Shresth and {de Miranda Cardoso}, Jos{'e} Vin{'i}cius and {Williams}, Peter K.~G. and {Wilson}, Tom J. and {Winkel}, Benjamin and {Wood-Vasey}, W.~M. and {Xue}, Rui and {Yoachim}, Peter and {Zhang}, Chen and {Zonca}, Andrea and {Astropy Project Contributors}},
        title = "{The Astropy Project: Sustaining and Growing a Community-oriented Open-source Project and the Latest Major Release (v5.0) of the Core Package}",
      journal = {\apj},
         year = 2022,
        month = aug,
       volume = {935},
       number = {2},
          eid = {167},
        pages = {167},
          doi = {10.3847/1538-4357/ac7c74},
archivePrefix = {arXiv},
       eprint = {2206.14220},
 primaryClass = {astro-ph.IM},
       adsurl = {https://ui.adsabs.harvard.edu/abs/2022ApJ...935..167A}
}

@ARTICLE{dm,
       author = {{G{\"o}ppl}, C. and {Preibisch}, T.},
        title = "{Gaia EDR3 distances of the young stellar clusters in the extended Carina Nebula complex}",
      journal = {\aap},
         year = 2022,
        month = apr,
       volume = {660},
          eid = {A11},
        pages = {A11},
          doi = {10.1051/0004-6361/202142576},
archivePrefix = {arXiv},
       eprint = {2201.09097},
 primaryClass = {astro-ph.SR},
       adsurl = {https://ui.adsabs.harvard.edu/abs/2022A&A...660A..11G}
}

@BOOK{mbol,
       author = {{Cox}, Arthur N.},
        title = "{Allen's astrophysical quantities}",
         year = 2000,
       adsurl = {https://ui.adsabs.harvard.edu/abs/2000asqu.book.....C}
}

@ARTICLE{proporion1,
       author = {{O'Dell}, C.~R. and {Wen}, Zheng and {Hu}, Xihai},
        title = "{Discovery of New Objects in the Orion Nebula on HST Images: Shocks, Compact Sources, and Protoplanetary Disks}",
      journal = {\apj},
         year = 1993,
        month = jun,
       volume = {410},
        pages = {696},
          doi = {10.1086/172786},
       adsurl = {https://ui.adsabs.harvard.edu/abs/1993ApJ...410..696O}
}

@ARTICLE{wh,
       author = {{Winter}, Andrew J. and {Haworth}, Thomas J.},
        title = "{The external photoevaporation of planet-forming discs}",
      journal = {European Physical Journal Plus},
         year = 2022,
        month = oct,
       volume = {137},
       number = {10},
          eid = {1132},
        pages = {1132},
          doi = {10.1140/epjp/s13360-022-03314-1},
archivePrefix = {arXiv},
       eprint = {2206.11910},
 primaryClass = {astro-ph.EP},
       adsurl = {https://ui.adsabs.harvard.edu/abs/2022EPJP..137.1132W}
}

@ARTICLE{otr14,
       author = {{Rainot}, A. and {Reggiani}, M. and {Sana}, H. and {Bodensteiner}, J. and {Absil}, O.},
        title = "{Carina High-contrast Imaging Project for massive Stars (CHIPS). II. O stars in Trumpler 14}",
      journal = {\aap},
         year = 2022,
        month = feb,
       volume = {658},
          eid = {A198},
        pages = {A198},
          doi = {10.1051/0004-6361/202141562},
archivePrefix = {arXiv},
       eprint = {2111.12361},
 primaryClass = {astro-ph.SR},
       adsurl = {https://ui.adsabs.harvard.edu/abs/2022A&A...658A.198R}
}

@ARTICLE{g0,
       author = {{Anania}, Rossella and {Winter}, Andrew J. and {Rosotti}, Giovanni and {Vioque}, Miguel and {Zari}, Eleonora and {Pantaleoni Gonz{\'a}lez}, Michelangelo and {Testi}, Leonardo},
        title = "{A novel method for estimating the far-ultraviolet flux, and a catalogue for disc-hosting stars in nearby star-forming regions}",
      journal = {\aap},
         year = 2025,
        month = mar,
       volume = {695},
          eid = {A74},
        pages = {A74},
          doi = {10.1051/0004-6361/202453011},
archivePrefix = {arXiv},
       eprint = {2501.18752},
 primaryClass = {astro-ph.EP},
       adsurl = {https://ui.adsabs.harvard.edu/abs/2025A&A...695A..74A}
}

@ARTICLE{veil_basri,
       author = {{Basri}, Gibor and {Batalha}, Celso},
        title = "{Hamilton Echelle Spectra of Young Stars. I. Optical Veiling}",
      journal = {\apj},
         year = 1990,
        month = nov,
       volume = {363},
        pages = {654},
          doi = {10.1086/169374},
       adsurl = {https://ui.adsabs.harvard.edu/abs/1990ApJ...363..654B}
}

@ARTICLE{veil_manara,
       author = {{Manara}, C.~F. and {Beccari}, G. and {Da Rio}, N. and {De Marchi}, G. and {Natta}, A. and {Ricci}, L. and {Robberto}, M. and {Testi}, L.},
        title = "{Accurate determination of accretion and photospheric parameters in young stellar objects: The case of two candidate old disks in the Orion Nebula Cluster}",
      journal = {\aap},
         year = 2013,
        month = oct,
       volume = {558},
          eid = {A114},
        pages = {A114},
          doi = {10.1051/0004-6361/201321866},
archivePrefix = {arXiv},
       eprint = {1307.8118},
 primaryClass = {astro-ph.SR},
       adsurl = {https://ui.adsabs.harvard.edu/abs/2013A&A...558A.114M}
}

@ARTICLE{veil_calvet,
       author = {{Calvet}, Nuria and {Gullbring}, Erik},
        title = "{The Structure and Emission of the Accretion Shock in T Tauri Stars}",
      journal = {\apj},
         year = 1998,
        month = dec,
       volume = {509},
       number = {2},
        pages = {802-818},
          doi = {10.1086/306527},
       adsurl = {https://ui.adsabs.harvard.edu/abs/1998ApJ...509..802C}
}

@ARTICLE{nir,
       author = {{Zeidler}, P. and {Preibisch}, T. and {Ratzka}, T. and {Roccatagliata}, V. and {Petr-Gotzens}, M.~G.},
        title = "{The VISTA Carina Nebula Survey. II. Spatial distribution of the infrared-excess-selected young stellar population}",
      journal = {\aap},
         year = 2016,
        month = jan,
       volume = {585},
          eid = {A49},
        pages = {A49},
          doi = {10.1051/0004-6361/201424376},
archivePrefix = {arXiv},
       eprint = {1510.01631},
 primaryClass = {astro-ph.SR},
       adsurl = {https://ui.adsabs.harvard.edu/abs/2016A&A...585A..49Z}
}

@ARTICLE{krumholz,
       author = {{Krumholz}, Mark R. and {McKee}, Christopher F. and {Bland-Hawthorn}, Joss},
        title = "{Star Clusters Across Cosmic Time}",
      journal = {\araa},
         year = 2019,
        month = aug,
       volume = {57},
        pages = {227-303},
          doi = {10.1146/annurev-astro-091918-104430},
archivePrefix = {arXiv},
       eprint = {1812.01615},
 primaryClass = {astro-ph.GA},
       adsurl = {https://ui.adsabs.harvard.edu/abs/2019ARA&A..57..227K}
}

@ARTICLE{ladalada,
       author = {{Lada}, Charles J. and {Lada}, Elizabeth A.},
        title = "{Embedded Clusters in Molecular Clouds}",
      journal = {\araa},
         year = 2003,
        month = jan,
       volume = {41},
        pages = {57-115},
          doi = {10.1146/annurev.astro.41.011802.094844},
archivePrefix = {arXiv},
       eprint = {astro-ph/0301540},
 primaryClass = {astro-ph},
       adsurl = {https://ui.adsabs.harvard.edu/abs/2003ARA&A..41...57L}
}

@ARTICLE{proporion2,
       author = {{Ricci}, L. and {Robberto}, M. and {Soderblom}, D.~R.},
        title = "{The Hubble Space Telescope/Advanced Camera for Surveys Atlas of Protoplanetary Disks in the Great Orion Nebula}",
      journal = {\aj},
         year = 2008,
        month = nov,
       volume = {136},
       number = {5},
        pages = {2136-2151},
          doi = {10.1088/0004-6256/136/5/2136},
       adsurl = {https://ui.adsabs.harvard.edu/abs/2008AJ....136.2136R}
}

@ARTICLE{ndugu,
       author = {{Ndugu}, N. and {Bitsch}, B. and {Jurua}, E.},
        title = "{Planet population synthesis driven by pebble accretion in cluster environments}",
      journal = {\mnras},
         year = 2018,
        month = feb,
       volume = {474},
       number = {1},
        pages = {886-897},
          doi = {10.1093/mnras/stx2815},
archivePrefix = {arXiv},
       eprint = {1710.10863},
 primaryClass = {astro-ph.EP},
       adsurl = {https://ui.adsabs.harvard.edu/abs/2018MNRAS.474..886N}
}

@ARTICLE{aru,
       author = {{Aru}, M.-L. and {Mauc{\'o}}, K. and {Manara}, C.~F. and {Haworth}, T.~J. and {Facchini}, S. and {McLeod}, A.~F. and {Miotello}, A. and {Petr-Gotzens}, M.~G. and {Robberto}, M. and {Rosotti}, G.~P. and {Vicente}, S. and {Winter}, A. and {Ansdell}, M.},
        title = "{Kaleidoscope of irradiated disks: MUSE observations of proplyds in the Orion Nebula Cluster. I. Sample presentation and ionization front sizes<xref rid=``FN2'' ref-type=``fn''/>}",
      journal = {\aap},
         year = 2024,
        month = jul,
       volume = {687},
          eid = {A93},
        pages = {A93},
          doi = {10.1051/0004-6361/202349004},
archivePrefix = {arXiv},
       eprint = {2403.12604},
 primaryClass = {astro-ph.SR},
       adsurl = {https://ui.adsabs.harvard.edu/abs/2024A&A...687A..93A}
}

@ARTICLE{winter2022,
       author = {{Winter}, Andrew J. and {Rosotti}, Giovanni P. and {Clarke}, Cathie and {Giersz}, Mirek},
        title = "{Forming short-period substellar companions in 47 Tucanae - I. Dynamical model and brown dwarf tidal capture rates}",
      journal = {\mnras},
         year = 2022,
        month = jan,
       volume = {509},
       number = {3},
        pages = {3924-3937},
          doi = {10.1093/mnras/stab3272},
archivePrefix = {arXiv},
       eprint = {2111.05372},
 primaryClass = {astro-ph.SR},
       adsurl = {https://ui.adsabs.harvard.edu/abs/2022MNRAS.509.3924W}
}

@ARTICLE{qiao2023,
       author = {{Qiao}, Lin and {Coleman}, Gavin A.~L. and {Haworth}, Thomas J.},
        title = "{Planet formation via pebble accretion in externally photoevaporating discs}",
      journal = {\mnras},
         year = 2023,
        month = jun,
       volume = {522},
       number = {2},
        pages = {1939-1950},
          doi = {10.1093/mnras/stad944},
archivePrefix = {arXiv},
       eprint = {2303.15177},
 primaryClass = {astro-ph.EP},
       adsurl = {https://ui.adsabs.harvard.edu/abs/2023MNRAS.522.1939Q}
}

@ARTICLE{huang2024a,
       author = {{Huang}, Shuo and {Portegies Zwart}, Simon and {Wilhelm}, Maite J.~C.},
        title = "{Suppression of giant planet formation around low-mass stars in clustered environments}",
      journal = {\aap},
         year = 2024,
        month = sep,
       volume = {689},
          eid = {A338},
        pages = {A338},
          doi = {10.1051/0004-6361/202451051},
archivePrefix = {arXiv},
       eprint = {2407.19018},
 primaryClass = {astro-ph.EP},
       adsurl = {https://ui.adsabs.harvard.edu/abs/2024A&A...689A.338H}
}

@ARTICLE{odell2017,
       author = {{O'Dell}, C.~R. and {Kollatschny}, W. and {Ferland}, G.~J.},
        title = "{Which Stars Are Ionizing the Orion Nebula?}",
      journal = {\apj},
         year = 2017,
        month = mar,
       volume = {837},
       number = {2},
          eid = {151},
        pages = {151},
          doi = {10.3847/1538-4357/aa6198},
       adsurl = {https://ui.adsabs.harvard.edu/abs/2017ApJ...837..151O}
}

@ARTICLE{harayama2008,
       author = {{Harayama}, Y. and {Eisenhauer}, F. and {Martins}, F.},
        title = "{The Initial Mass Function of the Massive Star-forming Region NGC 3603 from Near-Infrared Adaptive Optics Observations}",
      journal = {\apj},
         year = 2008,
        month = mar,
       volume = {675},
       number = {2},
        pages = {1319-1342},
          doi = {10.1086/524650},
archivePrefix = {arXiv},
       eprint = {0710.2882},
 primaryClass = {astro-ph},
       adsurl = {https://ui.adsabs.harvard.edu/abs/2008ApJ...675.1319H}
}

@ARTICLE{clark2005,
       author = {{Clark}, J.~S. and {Negueruela}, I. and {Crowther}, P.~A. and {Goodwin}, S.~P.},
        title = "{On the massive stellar population of the super star cluster <ASTROBJ>Westerlund 1</ASTROBJ>}",
      journal = {\aap},
         year = 2005,
        month = may,
       volume = {434},
       number = {3},
        pages = {949-969},
          doi = {10.1051/0004-6361:20042413},
archivePrefix = {arXiv},
       eprint = {astro-ph/0504342},
 primaryClass = {astro-ph},
       adsurl = {https://ui.adsabs.harvard.edu/abs/2005A&A...434..949C}
}

@ARTICLE{shull2021,
       author = {{Shull}, J. Michael and {Darling}, Jeremy and {Danforth}, Charles W.},
        title = "{Gaia EDR3 Parallax Distances to the Great Carina Nebula and Its Star Clusters (Trumpler 14, 15, 16)}",
      journal = {\apj},
         year = 2021,
        month = jun,
       volume = {914},
       number = {1},
          eid = {18},
        pages = {18},
          doi = {10.3847/1538-4357/abf4d8},
       adsurl = {https://ui.adsabs.harvard.edu/abs/2021ApJ...914...18S}
}

@INPROCEEDINGS{walborn1995,
       author = {{Walborn}, N.~R.},
        title = "{The Stellar Content of the Carina Nebula (Invited Paper)}",
    booktitle = {Revista Mexicana de Astronomia y Astrofisica Conference Series},
         year = 1995,
       editor = {{Niemela}, V. and {Morrell}, N. and {Feinstein}, A.},
       series = {Revista Mexicana de Astronomia y Astrofisica Conference Series},
       volume = {2},
        month = jun,
        pages = {51},
       adsurl = {https://ui.adsabs.harvard.edu/abs/1995RMxAC...2...51W}
}

@ARTICLE{hur2023,
       author = {{Hur}, Hyeonoh and {Lim}, Beomdu and {Chun}, Moo-Young},
        title = "{A Deep Optical Survey of Young Stars in the Carina Nebula. I. UBVRI Photometric Data and Fundamental Parameters}",
      journal = {Journal of Korean Astronomical Society},
         year = 2023,
        month = may,
       volume = {56},
        pages = {97-115},
          doi = {10.5303/JKAS.2023.56.1.97},
archivePrefix = {arXiv},
       eprint = {2305.01887},
 primaryClass = {astro-ph.SR},
       adsurl = {https://ui.adsabs.harvard.edu/abs/2023JKAS...56...97H}
}

@ARTICLE{berlanas2023,
       author = {{Berlanas}, S.~R. and {Ma{\'\i}z Apell{\'a}niz}, J. and {Herrero}, A. and {Mahy}, L. and {Blomme}, R. and {Negueruela}, I. and {Dorda}, R. and {Comer{\'o}n}, F. and {Gosset}, E. and {Pantaleoni Gonz{\'a}lez}, M. and {Molina Lera}, J.~A. and {Sota}, A. and {Furst}, T. and {Alfaro}, E.~J. and {Bergemann}, M. and {Carraro}, G. and {Drew}, J.~E. and {Morbidelli}, L. and {Vink}, J.~S.},
        title = "{Gaia-ESO survey: Massive stars in the Carina Nebula. I. A new census of OB stars}",
      journal = {\aap},
         year = 2023,
        month = mar,
       volume = {671},
          eid = {A20},
        pages = {A20},
          doi = {10.1051/0004-6361/202245335},
archivePrefix = {arXiv},
       eprint = {2301.08310},
 primaryClass = {astro-ph.SR},
       adsurl = {https://ui.adsabs.harvard.edu/abs/2023A&A...671A..20B}
}

@ARTICLE{odell1994,
       author = {{O'Dell}, C.~R. and {Wen}, Zheng},
        title = "{Postrefurbishment Mission Hubble Space Telescope Images of the Core of the Orion Nebula: Proplyds, Herbig-Haro Objects, and Measurements of a Circumstellar Disk}",
      journal = {\apj},
         year = 1994,
        month = nov,
       volume = {436},
        pages = {194},
          doi = {10.1086/174892},
       adsurl = {https://ui.adsabs.harvard.edu/abs/1994ApJ...436..194O}
}

@ARTICLE{mccaugh1996,
       author = {{McCaughrean}, Mark J. and {O'Dell}, C. Robert},
        title = "{Direct Imaging of Circumstellar Disks in the Orion Nebula}",
      journal = {\aj},
         year = 1996,
        month = may,
       volume = {111},
        pages = {1977},
          doi = {10.1086/117934},
       adsurl = {https://ui.adsabs.harvard.edu/abs/1996AJ....111.1977M}
}

@ARTICLE{penny1993,
       author = {{Penny}, Laura R. and {Gies}, Douglas R. and {Hartkopf}, William I. and {Mason}, Brian D. and {Turner}, Nils H.},
        title = "{The Frequency of Binary Stars in the Young Cluster Trumpler 14}",
      journal = {\pasp},
         year = 1993,
        month = jun,
       volume = {105},
        pages = {588},
          doi = {10.1086/133200},
       adsurl = {https://ui.adsabs.harvard.edu/abs/1993PASP..105..588P}
}

@ARTICLE{vasquez1996,
       author = {{Vazquez}, R.~A. and {Baume}, G. and {Feinstein}, A. and {Prado}, P.},
        title = "{Investigation on the region of the open cluster Tr 14.}",
      journal = {\aaps},
         year = 1996,
        month = mar,
       volume = {116},
        pages = {75-94},
       adsurl = {https://ui.adsabs.harvard.edu/abs/1996A&AS..116...75V}
}

@ARTICLE{degioia2001,
       author = {{DeGioia-Eastwood}, K. and {Throop}, H. and {Walker}, G. and {Cudworth}, K.~M.},
        title = "{The Star Formation History of Trumpler 14 and Trumpler 16}",
      journal = {\apj},
         year = 2001,
        month = mar,
       volume = {549},
       number = {1},
        pages = {578-589},
          doi = {10.1086/319047},
       adsurl = {https://ui.adsabs.harvard.edu/abs/2001ApJ...549..578D}
}

@ARTICLE{carraro2004,
       author = {{Carraro}, G. and {Romaniello}, M. and {Ventura}, P. and {Patat}, F.},
        title = "{The star cluster Collinder 232 in the Carina complex and its relation to Trumpler 14/16}",
      journal = {\aap},
         year = 2004,
        month = may,
       volume = {418},
        pages = {525-537},
          doi = {10.1051/0004-6361:20034335},
archivePrefix = {arXiv},
       eprint = {astro-ph/0401144},
 primaryClass = {astro-ph},
       adsurl = {https://ui.adsabs.harvard.edu/abs/2004A&A...418..525C}
}

@ARTICLE{haworth2018a,
       author = {{Haworth}, Thomas J. and {Clarke}, Cathie J. and {Rahman}, Wahidur and {Winter}, Andrew J. and {Facchini}, Stefano},
        title = "{The FRIED grid of mass-loss rates for externally irradiated protoplanetary discs}",
      journal = {\mnras},
         year = 2018,
        month = nov,
       volume = {481},
       number = {1},
        pages = {452-466},
          doi = {10.1093/mnras/sty2323},
archivePrefix = {arXiv},
       eprint = {1808.07484},
 primaryClass = {astro-ph.SR},
       adsurl = {https://ui.adsabs.harvard.edu/abs/2018MNRAS.481..452H}
}

@ARTICLE{manara2013,
       author = {{Manara}, C.~F. and {Testi}, L. and {Rigliaco}, E. and {Alcal{\'a}}, J.~M. and {Natta}, A. and {Stelzer}, B. and {Biazzo}, K. and {Covino}, E. and {Covino}, S. and {Cupani}, G. and {D'Elia}, V. and {Randich}, S.},
        title = "{X-shooter spectroscopy of young stellar objects. II. Impact of chromospheric emission on accretion rate estimates}",
      journal = {\aap},
         year = 2013,
        month = mar,
       volume = {551},
          eid = {A107},
        pages = {A107},
          doi = {10.1051/0004-6361/201220921},
archivePrefix = {arXiv},
       eprint = {1301.3058},
 primaryClass = {astro-ph.GA},
       adsurl = {https://ui.adsabs.harvard.edu/abs/2013A&A...551A.107M}
}

@ARTICLE{manara2017,
       author = {{Manara}, C.~F. and {Frasca}, A. and {Alcal{\'a}}, J.~M. and {Natta}, A. and {Stelzer}, B. and {Testi}, L.},
        title = "{An extensive VLT/X-shooter library of photospheric templates of pre-main sequence stars}",
      journal = {\aap},
         year = 2017,
        month = sep,
       volume = {605},
          eid = {A86},
        pages = {A86},
          doi = {10.1051/0004-6361/201730807},
archivePrefix = {arXiv},
       eprint = {1705.10075},
 primaryClass = {astro-ph.SR},
       adsurl = {https://ui.adsabs.harvard.edu/abs/2017A&A...605A..86M}
}

@ARTICLE{ardrizzone2019b,
       author = {{Ardizzone}, Lynton and {L{\"u}th}, Carsten and {Kruse}, Jakob and {Rother}, Carsten and {K{\"o}the}, Ullrich},
        title = "{Guided Image Generation with Conditional Invertible Neural Networks}",
      journal = {arXiv e-prints},
         year = 2019,
        month = jul,
          eid = {arXiv:1907.02392},
        pages = {arXiv:1907.02392},
          doi = {10.48550/arXiv.1907.02392},
archivePrefix = {arXiv},
       eprint = {1907.02392},
 primaryClass = {cs.CV},
       adsurl = {https://ui.adsabs.harvard.edu/abs/2019arXiv190702392A}
}

@ARTICLE{chabrier,
       author = {{Chabrier}, Gilles},
        title = "{Galactic Stellar and Substellar Initial Mass Function}",
      journal = {\pasp},
         year = 2003,
        month = jul,
       volume = {115},
       number = {809},
        pages = {763-795},
          doi = {10.1086/376392},
archivePrefix = {arXiv},
       eprint = {astro-ph/0304382},
 primaryClass = {astro-ph},
       adsurl = {https://ui.adsabs.harvard.edu/abs/2003PASP..115..763C}
}

@ARTICLE{kroupa,
       author = {{Kroupa}, Pavel},
        title = "{On the variation of the initial mass function}",
      journal = {\mnras},
         year = 2001,
        month = apr,
       volume = {322},
       number = {2},
        pages = {231-246},
          doi = {10.1046/j.1365-8711.2001.04022.x},
archivePrefix = {arXiv},
       eprint = {astro-ph/0009005},
 primaryClass = {astro-ph},
       adsurl = {https://ui.adsabs.harvard.edu/abs/2001MNRAS.322..231K}
}

@ARTICLE{habing,
       author = {{Habing}, H.~J.},
        title = "{The interstellar radiation density between 912 A and 2400 A}",
      journal = {\bain},
         year = 1968,
        month = jan,
       volume = {19},
        pages = {421},
       adsurl = {https://ui.adsabs.harvard.edu/abs/1968BAN....19..421H}
}

@ARTICLE{guarcello,
       author = {{Guarcello}, M.~G. and {Drake}, J.~J. and {Wright}, N.~J. and {Albacete-Colombo}, J.~F. and {Clarke}, C. and {Ercolano}, B. and {Flaccomio}, E. and {Kashyap}, V. and {Micela}, G. and {Naylor}, T. and {Schneider}, N. and {Sciortino}, S. and {Vink}, J.~S.},
        title = "{Photoevaporation and Close Encounters: How the Environment around Cygnus OB2 Affects the Evolution of Protoplanetary Disks}",
      journal = {\apjs},
         year = 2023,
        month = nov,
       volume = {269},
       number = {1},
          eid = {13},
        pages = {13},
          doi = {10.3847/1538-4365/acdd67},
       adsurl = {https://ui.adsabs.harvard.edu/abs/2023ApJS..269...13G}
}

@ARTICLE{ascenso,
       author = {{Ascenso}, J. and {Alves}, J. and {Vicente}, S. and {Lago}, M.~T.~V.~T.},
        title = "{NTT and VLT diffraction limited imaging of Trumpler 14: revealing a massive core-halo cluster}",
      journal = {\aap},
         year = 2007,
        month = dec,
       volume = {476},
       number = {1},
        pages = {199-215},
          doi = {10.1051/0004-6361:20077210},
       adsurl = {https://ui.adsabs.harvard.edu/abs/2007A&A...476..199A}
}

@ARTICLE{guarc2007,
       author = {{Guarcello}, M.~G. and {Prisinzano}, L. and {Micela}, G. and {Damiani}, F. and {Peres}, G. and {Sciortino}, S.},
        title = "{Correlation between the spatial distribution of circumstellar disks and massive stars in the open cluster NGC 6611. Compiled catalog and cluster parameters}",
      journal = {\aap},
         year = 2007,
        month = jan,
       volume = {462},
       number = {1},
        pages = {245-255},
          doi = {10.1051/0004-6361:20066124},
archivePrefix = {arXiv},
       eprint = {astro-ph/0610401},
 primaryClass = {astro-ph},
       adsurl = {https://ui.adsabs.harvard.edu/abs/2007A&A...462..245G}
}

@ARTICLE{guarc2009,
       author = {{Guarcello}, M.~G. and {Micela}, G. and {Damiani}, F. and {Peres}, G. and {Prisinzano}, L. and {Sciortino}, S.},
        title = "{Correlation between the spatial distribution of circumstellar disks and massive stars in the young open cluster NGC 6611. II. Cluster members selected with Spitzer/IRAC}",
      journal = {\aap},
         year = 2009,
        month = mar,
       volume = {496},
       number = {2},
        pages = {453-463},
          doi = {10.1051/0004-6361/200810671},
archivePrefix = {arXiv},
       eprint = {0812.0945},
 primaryClass = {astro-ph},
       adsurl = {https://ui.adsabs.harvard.edu/abs/2009A&A...496..453G}
}

@ARTICLE{guarc2010,
       author = {{Guarcello}, M.~G. and {Micela}, G. and {Peres}, G. and {Prisinzano}, L. and {Sciortino}, S.},
        title = "{Chronology of star formation and disk evolution in the Eagle Nebula}",
      journal = {\aap},
         year = 2010,
        month = oct,
       volume = {521},
          eid = {A61},
        pages = {A61},
          doi = {10.1051/0004-6361/201014351},
archivePrefix = {arXiv},
       eprint = {1008.0422},
 primaryClass = {astro-ph.SR},
       adsurl = {https://ui.adsabs.harvard.edu/abs/2010A&A...521A..61G}
}

@ARTICLE{rygl,
       author = {{Rygl}, K.~L.~J. and {Brunthaler}, A. and {Sanna}, A. and {Menten}, K.~M. and {Reid}, M.~J. and {van Langevelde}, H.~J. and {Honma}, M. and {Torstensson}, K.~J.~E. and {Fujisawa}, K.},
        title = "{Parallaxes and proper motions of interstellar masers toward the Cygnus X star-forming complex. I. Membership of the Cygnus X region}",
      journal = {\aap},
         year = 2012,
        month = mar,
       volume = {539},
          eid = {A79},
        pages = {A79},
          doi = {10.1051/0004-6361/201118211},
archivePrefix = {arXiv},
       eprint = {1111.7023},
 primaryClass = {astro-ph.GA},
       adsurl = {https://ui.adsabs.harvard.edu/abs/2012A&A...539A..79R}
}

@ARTICLE{wright2010,
       author = {{Wright}, N.~J. and {Drake}, J.~J. and {Drew}, J.~E. and {Vink}, J.~S.},
        title = "{The Massive Star-Forming Region Cygnus OB2. II. Integrated Stellar Properties and the Star Formation History}",
      journal = {\apj},
         year = 2010,
        month = apr,
       volume = {713},
       number = {2},
        pages = {871-882},
          doi = {10.1088/0004-637X/713/2/871},
archivePrefix = {arXiv},
       eprint = {1003.2463},
 primaryClass = {astro-ph.SR},
       adsurl = {https://ui.adsabs.harvard.edu/abs/2010ApJ...713..871W}
}

@ARTICLE{hanson2003,
       author = {{Hanson}, M.~M.},
        title = "{A Study of Cygnus OB2: Pointing the Way toward Finding Our Galaxy's Super-Star Clusters}",
      journal = {\apj},
         year = 2003,
        month = nov,
       volume = {597},
       number = {2},
        pages = {957-969},
          doi = {10.1086/378508},
archivePrefix = {arXiv},
       eprint = {astro-ph/0307540},
 primaryClass = {astro-ph},
       adsurl = {https://ui.adsabs.harvard.edu/abs/2003ApJ...597..957H}
}

@ARTICLE{drew2008,
       author = {{Drew}, Janet E. and {Greimel}, R. and {Irwin}, M.~J. and {Sale}, S.~E.},
        title = "{Early-A stars from IPHAS, and their distribution in and around the Cyg OB2 association}",
      journal = {\mnras},
         year = 2008,
        month = jun,
       volume = {386},
       number = {4},
        pages = {1761-1773},
          doi = {10.1111/j.1365-2966.2008.13147.x},
archivePrefix = {arXiv},
       eprint = {0802.3868},
 primaryClass = {astro-ph},
       adsurl = {https://ui.adsabs.harvard.edu/abs/2008MNRAS.386.1761D}
}

@ARTICLE{guarc2013,
       author = {{Guarcello}, M.~G. and {Drake}, J.~J. and {Wright}, N.~J. and {Drew}, J.~E. and {Gutermuth}, R.~A. and {Hora}, J.~L. and {Naylor}, T. and {Aldcroft}, T. and {Fruscione}, A. and {Garc{\'\i}a-Alvarez}, D. and {Kashyap}, V.~L. and {King}, R.},
        title = "{The Protoplanetary Disks in the Nearby Massive Star-forming Region Cygnus OB2}",
      journal = {\apj},
         year = 2013,
        month = aug,
       volume = {773},
       number = {2},
          eid = {135},
        pages = {135},
          doi = {10.1088/0004-637X/773/2/135},
archivePrefix = {arXiv},
       eprint = {1306.5757},
 primaryClass = {astro-ph.SR},
       adsurl = {https://ui.adsabs.harvard.edu/abs/2013ApJ...773..135G}
}

@ARTICLE{qiao2026,
       author = {{Qiao}, Lin and {Coleman}, Gavin A.~L. and {Haworth}, Thomas J.},
        title = "{Formation of multiplanetary systems via pebble accretion in externally photoevaporating discs in stellar clusters}",
      journal = {\mnras},
         year = 2026,
        month = mar,
       volume = {546},
       number = {3},
          eid = {stag034},
        pages = {stag034},
          doi = {10.1093/mnras/stag034},
archivePrefix = {arXiv},
       eprint = {2601.03963},
 primaryClass = {astro-ph.EP},
       adsurl = {https://ui.adsabs.harvard.edu/abs/2026MNRAS.546ag034Q}
}

@ARTICLE{megeath2012,
       author = {{Megeath}, S.~T. and {Gutermuth}, R. and {Muzerolle}, J. and {Kryukova}, E. and {Flaherty}, K. and {Hora}, J.~L. and {Allen}, L.~E. and {Hartmann}, L. and {Myers}, P.~C. and {Pipher}, J.~L. and {Stauffer}, J. and {Young}, E.~T. and {Fazio}, G.~G.},
        title = "{The Spitzer Space Telescope Survey of the Orion A and B Molecular Clouds. I. A Census of Dusty Young Stellar Objects and a Study of Their Mid-infrared Variability}",
      journal = {\aj},
         year = 2012,
        month = dec,
       volume = {144},
       number = {6},
          eid = {192},
        pages = {192},
          doi = {10.1088/0004-6256/144/6/192},
archivePrefix = {arXiv},
       eprint = {1209.3826},
 primaryClass = {astro-ph.GA},
       adsurl = {https://ui.adsabs.harvard.edu/abs/2012AJ....144..192M}
}

@ARTICLE{wright2020,
       author = {{Wright}, Nicholas J.},
        title = "{OB Associations and their origins}",
      journal = {\nar},
         year = 2020,
        month = nov,
       volume = {90},
          eid = {101549},
        pages = {101549},
          doi = {10.1016/j.newar.2020.101549},
archivePrefix = {arXiv},
       eprint = {2011.09483},
 primaryClass = {astro-ph.SR},
       adsurl = {https://ui.adsabs.harvard.edu/abs/2020NewAR..9001549W}
}

@ARTICLE{negeruela2022,
       author = {{Negueruela}, I. and {Alfaro}, E.~J. and {Dorda}, R. and {Marco}, A. and {Ma{\'\i}z Apell{\'a}niz}, J. and {Gonz{\'a}lez-Fern{\'a}ndez}, C.},
        title = "{Westerlund 1 under the light of Gaia EDR3: Distance, isolation, extent, and a hidden population}",
      journal = {\aap},
         year = 2022,
        month = aug,
       volume = {664},
          eid = {A146},
        pages = {A146},
          doi = {10.1051/0004-6361/202142985},
archivePrefix = {arXiv},
       eprint = {2204.00422},
 primaryClass = {astro-ph.SR},
       adsurl = {https://ui.adsabs.harvard.edu/abs/2022A&A...664A.146N}
}

@ARTICLE{damiani2017,
       author = {{Damiani}, F. and {Klutsch}, A. and {Jeffries}, R.~D. and {Randich}, S. and {Prisinzano}, L. and {Ma{\'\i}z Apell{\'a}niz}, J. and {Micela}, G. and {Kalari}, V. and {Frasca}, A. and {Zwitter}, T. and {Bonito}, R. and {Gilmore}, G. and {Flaccomio}, E. and {Francois}, P. and {Koposov}, S. and {Lanzafame}, A.~C. and {Sacco}, G.~G. and {Bayo}, A. and {Carraro}, G. and {Casey}, A.~R. and {Alfaro}, E.~J. and {Costado}, M.~T. and {Donati}, P. and {Franciosini}, E. and {Hourihane}, A. and {Jofr{\'e}}, P. and {Lardo}, C. and {Lewis}, J. and {Magrini}, L. and {Monaco}, L. and {Morbidelli}, L. and {Worley}, C.~C. and {Vink}, J.~S. and {Zaggia}, S.},
        title = "{Gaia-ESO Survey: Global properties of clusters Trumpler 14 and 16 in the Carina nebula}",
      journal = {\aap},
         year = 2017,
        month = jul,
       volume = {603},
          eid = {A81},
        pages = {A81},
          doi = {10.1051/0004-6361/201629020},
archivePrefix = {arXiv},
       eprint = {1702.04776},
 primaryClass = {astro-ph.SR},
       adsurl = {https://ui.adsabs.harvard.edu/abs/2017A&A...603A..81D}
}

@ARTICLE{goppl2022,
       author = {{G{\"o}ppl}, C. and {Preibisch}, T.},
        title = "{Gaia EDR3 distances of the young stellar clusters in the extended Carina Nebula complex}",
      journal = {\aap},
         year = 2022,
        month = apr,
       volume = {660},
          eid = {A11},
        pages = {A11},
          doi = {10.1051/0004-6361/202142576},
archivePrefix = {arXiv},
       eprint = {2201.09097},
 primaryClass = {astro-ph.SR},
       adsurl = {https://ui.adsabs.harvard.edu/abs/2022A&A...660A..11G}
}

@ARTICLE{veil_herczeg,
       author = {{Herczeg}, Gregory J. and {Hillenbrand}, Lynne A.},
        title = "{An Optical Spectroscopic Study of T Tauri Stars. I. Photospheric Properties}",
      journal = {\apj},
         year = 2014,
        month = may,
       volume = {786},
       number = {2},
          eid = {97},
        pages = {97},
          doi = {10.1088/0004-637X/786/2/97},
archivePrefix = {arXiv},
       eprint = {1403.1675},
 primaryClass = {astro-ph.SR},
       adsurl = {https://ui.adsabs.harvard.edu/abs/2014ApJ...786...97H}
}

@ARTICLE{parker,
       author = {{Parker}, Richard J.},
        title = "{The birth environment of planetary systems}",
      journal = {Royal Society Open Science},
         year = 2020,
        month = nov,
       volume = {7},
       number = {11},
          eid = {201271},
        pages = {201271},
          doi = {10.1098/rsos.201271},
archivePrefix = {arXiv},
       eprint = {2007.07890},
 primaryClass = {astro-ph.EP},
       adsurl = {https://ui.adsabs.harvard.edu/abs/2020RSOS....701271P}
}

@ARTICLE{weder2026,
       author = {{Weder}, Jesse and {Winter}, Andrew J. and {Mordasini}, Christoph},
        title = "{Inferring the physics of protoplanetary disc evolution from the irradiated Cygnus OB2 region: A comparison of viscous and MHD wind-driven scenarios}",
      journal = {\aap},
         year = 2026,
        month = jan,
       volume = {705},
          eid = {A102},
        pages = {A102},
          doi = {10.1051/0004-6361/202556316},
archivePrefix = {arXiv},
       eprint = {2511.01972},
 primaryClass = {astro-ph.EP},
       adsurl = {https://ui.adsabs.harvard.edu/abs/2026A&A...705A.102W}
}

@ARTICLE{sota2014,
       author = {{Sota}, A. and {Ma{\'\i}z Apell{\'a}niz}, J. and {Morrell}, N.~I. and {Barb{\'a}}, R.~H. and {Walborn}, N.~R. and {Gamen}, R.~C. and {Arias}, J.~I. and {Alfaro}, E.~J.},
        title = "{The Galactic O-Star Spectroscopic Survey (GOSSS). II. Bright Southern Stars}",
      journal = {\apjs},
         year = 2014,
        month = mar,
       volume = {211},
       number = {1},
          eid = {10},
        pages = {10},
          doi = {10.1088/0067-0049/211/1/10},
archivePrefix = {arXiv},
       eprint = {1312.6222},
 primaryClass = {astro-ph.GA},
       adsurl = {https://ui.adsabs.harvard.edu/abs/2014ApJS..211...10S}
}

@ARTICLE{gruner2019,
       author = {{Gruner}, D. and {Hainich}, R. and {Sander}, A.~A.~C. and {Shenar}, T. and {Todt}, H. and {Oskinova}, L.~M. and {Ramachandran}, V. and {Ayres}, T. and {Hamann}, W.-R.},
        title = "{The extreme O-type spectroscopic binary HD 93129A. A quantitative, multiwavelength analysis}",
      journal = {\aap},
         year = 2019,
        month = jan,
       volume = {621},
          eid = {A63},
        pages = {A63},
          doi = {10.1051/0004-6361/201833178},
archivePrefix = {arXiv},
       eprint = {1811.07820},
 primaryClass = {astro-ph.SR},
       adsurl = {https://ui.adsabs.harvard.edu/abs/2019A&A...621A..63G}
}

@INPROCEEDINGS{muse,
       author = {{Bacon}, R. and {Accardo}, M. and {Adjali}, L. and {Anwand}, H. and {Bauer}, S. and {Biswas}, I. and {Blaizot}, J. and {Boudon}, D. and {Brau-Nogue}, S. and {Brinchmann}, J. and {Caillier}, P. and {Capoani}, L. and {Carollo}, C.~M. and {Contini}, T. and {Couderc}, P. and {Daguis{\'e}}, E. and {Deiries}, S. and {Delabre}, B. and {Dreizler}, S. and {Dubois}, J. and {Dupieux}, M. and {Dupuy}, C. and {Emsellem}, E. and {Fechner}, T. and {Fleischmann}, A. and {Fran{\c{c}}ois}, M. and {Gallou}, G. and {Gharsa}, T. and {Glindemann}, A. and {Gojak}, D. and {Guiderdoni}, B. and {Hansali}, G. and {Hahn}, T. and {Jarno}, A. and {Kelz}, A. and {Koehler}, C. and {Kosmalski}, J. and {Laurent}, F. and {Le Floch}, M. and {Lilly}, S.~J. and {Lizon}, J.-L. and {Loupias}, M. and {Manescau}, A. and {Monstein}, C. and {Nicklas}, H. and {Olaya}, J.-C. and {Pares}, L. and {Pasquini}, L. and {P{\'e}contal-Rousset}, A. and {Pell{\'o}}, R. and {Petit}, C. and {Popow}, E. and {Reiss}, R. and {Remillieux}, A. and {Renault}, E. and {Roth}, M. and {Rupprecht}, G. and {Serre}, D. and {Schaye}, J. and {Soucail}, G. and {Steinmetz}, M. and {Streicher}, O. and {Stuik}, R. and {Valentin}, H. and {Vernet}, J. and {Weilbacher}, P. and {Wisotzki}, L. and {Yerle}, N.},
        title = "{The MUSE second-generation VLT instrument}",
    booktitle = {Ground-based and Airborne Instrumentation for Astronomy III},
         year = 2010,
       editor = {{McLean}, Ian S. and {Ramsay}, Suzanne K. and {Takami}, Hideki},
       series = {Society of Photo-Optical Instrumentation Engineers (SPIE) Conference Series},
       volume = {7735},
        month = jul,
          eid = {773508},
        pages = {773508},
          doi = {10.1117/12.856027},
archivePrefix = {arXiv},
       eprint = {2211.16795},
 primaryClass = {astro-ph.IM},
       adsurl = {https://ui.adsabs.harvard.edu/abs/2010SPIE.7735E..08B}
}

@ARTICLE{lorenzetti2013,
       author = {{Lorenzetti}, D. and {Antoniucci}, S. and {Giannini}, T. and {Di Paola}, A. and {Arkharov}, A.~A. and {Larionov}, V.~M.},
        title = "{Interpreting the simultaneous variability of near-IR continuum and line emission in young stellar objects}",
      journal = {\apss},
         year = 2013,
        month = feb,
       volume = {343},
       number = {2},
        pages = {535-539},
          doi = {10.1007/s10509-012-1266-4},
archivePrefix = {arXiv},
       eprint = {1210.0317},
 primaryClass = {astro-ph.SR},
       adsurl = {https://ui.adsabs.harvard.edu/abs/2013Ap&SS.343..535L}
}

@ARTICLE{kim2016,
       author = {{Kim}, Jinyoung Serena and {Clarke}, Cathie J. and {Fang}, Min and {Facchini}, Stefano},
        title = "{Proplyds Around a B1 Star: 42 Orionis in NGC 1977}",
      journal = {\apjl},
         year = 2016,
        month = jul,
       volume = {826},
       number = {1},
          eid = {L15},
        pages = {L15},
          doi = {10.3847/2041-8205/826/1/L15},
archivePrefix = {arXiv},
       eprint = {1606.08271},
 primaryClass = {astro-ph.SR},
       adsurl = {https://ui.adsabs.harvard.edu/abs/2016ApJ...826L..15K}
}

@ARTICLE{haworth2021,
       author = {{Haworth}, Thomas J. and {Kim}, Jinyoung S. and {Winter}, Andrew J. and {Hines}, Dean C. and {Clarke}, Cathie J. and {Sellek}, Andrew D. and {Ballabio}, Giulia and {Stapelfeldt}, Karl R.},
        title = "{Proplyds in the flame nebula NGC 2024}",
      journal = {\mnras},
         year = 2021,
        month = mar,
       volume = {501},
       number = {3},
        pages = {3502-3514},
          doi = {10.1093/mnras/staa3918},
archivePrefix = {arXiv},
       eprint = {2012.09166},
 primaryClass = {astro-ph.SR},
       adsurl = {https://ui.adsabs.harvard.edu/abs/2021MNRAS.501.3502H}
}

@ARTICLE{mesa2016,
       author = {{Mesa-Delgado}, A. and {Zapata}, L. and {Henney}, W.~J. and {Puzia}, T.~H. and {Tsamis}, Y.~G.},
        title = "{Protoplanetary Disks in the Hostile Environment of Carina}",
      journal = {\apjl},
         year = 2016,
        month = jul,
       volume = {825},
       number = {1},
          eid = {L16},
        pages = {L16},
          doi = {10.3847/2041-8205/825/1/L16},
archivePrefix = {arXiv},
       eprint = {1605.08809},
 primaryClass = {astro-ph.EP},
       adsurl = {https://ui.adsabs.harvard.edu/abs/2016ApJ...825L..16M}
}

@ARTICLE{clarke2007,
       author = {{Clarke}, C.~J.},
        title = "{The photoevaporation of discs around young stars in massive clusters}",
      journal = {\mnras},
         year = 2007,
        month = apr,
       volume = {376},
       number = {3},
        pages = {1350-1356},
          doi = {10.1111/j.1365-2966.2007.11547.x},
archivePrefix = {arXiv},
       eprint = {astro-ph/0702112},
 primaryClass = {astro-ph},
       adsurl = {https://ui.adsabs.harvard.edu/abs/2007MNRAS.376.1350C}
}

@ARTICLE{coleman2022,
       author = {{Coleman}, Gavin A.~L. and {Haworth}, Thomas J.},
        title = "{Dispersal of protoplanetary discs: how stellar properties and the local environment determine the pathway of evolution}",
      journal = {\mnras},
         year = 2022,
        month = aug,
       volume = {514},
       number = {2},
        pages = {2315-2332},
          doi = {10.1093/mnras/stac1513},
archivePrefix = {arXiv},
       eprint = {2204.02303},
 primaryClass = {astro-ph.EP},
       adsurl = {https://ui.adsabs.harvard.edu/abs/2022MNRAS.514.2315C}
}

@ARTICLE{concha2019,
       author = {{Concha-Ram{\'\i}rez}, Francisca and {Wilhelm}, Maite J.~C. and {Portegies Zwart}, Simon and {Haworth}, Thomas J.},
        title = "{External photoevaporation of circumstellar discs constrains the time-scale for planet formation}",
      journal = {\mnras},
         year = 2019,
        month = oct,
       volume = {490},
       number = {4},
        pages = {5678-5690},
          doi = {10.1093/mnras/stz2973},
archivePrefix = {arXiv},
       eprint = {1907.03760},
 primaryClass = {astro-ph.EP},
       adsurl = {https://ui.adsabs.harvard.edu/abs/2019MNRAS.490.5678C}
}

@ARTICLE{williams2011,
       author = {{Williams}, Jonathan P. and {Cieza}, Lucas A.},
        title = "{Protoplanetary Disks and Their Evolution}",
      journal = {\araa},
         year = 2011,
        month = sep,
       volume = {49},
       number = {1},
        pages = {67-117},
          doi = {10.1146/annurev-astro-081710-102548},
archivePrefix = {arXiv},
       eprint = {1103.0556},
 primaryClass = {astro-ph.GA},
       adsurl = {https://ui.adsabs.harvard.edu/abs/2011ARA&A..49...67W}
}

@ARTICLE{johnstone1998,
       author = {{Johnstone}, Doug and {Hollenbach}, David and {Bally}, John},
        title = "{Photoevaporation of Disks and Clumps by Nearby Massive Stars: Application to Disk Destruction in the Orion Nebula}",
      journal = {\apj},
         year = 1998,
        month = may,
       volume = {499},
       number = {2},
        pages = {758-776},
          doi = {10.1086/305658},
       adsurl = {https://ui.adsabs.harvard.edu/abs/1998ApJ...499..758J}
}

@ARTICLE{scally2001,
       author = {{Scally}, Aylwyn and {Clarke}, Cathie},
        title = "{Destruction of protoplanetary discs in the Orion Nebula Cluster}",
      journal = {\mnras},
         year = 2001,
        month = aug,
       volume = {325},
       number = {2},
        pages = {449-456},
          doi = {10.1046/j.1365-8711.2001.04274.x},
archivePrefix = {arXiv},
       eprint = {astro-ph/0012098},
 primaryClass = {astro-ph},
       adsurl = {https://ui.adsabs.harvard.edu/abs/2001MNRAS.325..449S}
}

@ARTICLE{cinn3,
       author = {{Kang}, D.E. and {Ksoll}, V. F.},
        title = "{Spectral classification of young stars using conditional invertible neural networks. III. SAPSAL: A universal, instrument-independent framework for stellar and accretion analysis}",
      journal = {\aap},
    year = {2026},
    note = {submitted}
}

\begin{appendix}
\section{NIR photometrical errors \label{app:nir_err}}

In Sect.~\ref{sec:nirex}, we study the disk fraction in the center of Tr14 using the NIR excess definition of \citet{nir}. We find 17\% of our sample are NIR excess sources, based on their position in the $(J-H)$ vs. $(H-K_s)$ diagram. While \citet{nir} consider a classification line shifted to incorporate an average error of 0.05 mag, if we want to examine in detail the photometric errors and their impact to the selection of the NIR excess sources, there are three cases to study:

\begin{enumerate}
    \item Following \citet{preibis1}, considering an average photometric error of 0.05 mag.
    \item When comparing with the 2MASS survey, \citet{preibis1} get the uncertainties: $\sigma_{J-H}=0.12\text{ mag}$, $\sigma_{H-K_s}=0.11\text{ mag}$.
    \item Following the HAWK-I manual, the individual photometric uncertainties suggested are 0.05 mag. This yields a color-color uncertainty of 0.071 mag. 
\end{enumerate}
                
To quantify the effect of these photometric errors on the computed disk fractions, we adopt a probabilistic approach. We reconstruct the $(J-H)$ vs. $(H-K_s)$ diagram as a two-dimensional density map on a regular pixel grid. The color-color space is mapped onto an image matrix of $N_{(H-K_s)}\times N_{(J-H)}$ pixels spanning the ranges of $(H-K_s)$ and $(J-H)$ accordingly. Each source is represented by a normalized two-dimensional Gaussian kernel centered on its measured position in the color-difference planes. The kernel widths are determined by the corresponding color uncertainties $\sigma_{H-K_s},\ \sigma_{J-H}$. The final density map is obtained by summing the contribution of all individual kernels. 

To estimate the fraction of sources exhibiting NIR excess emission, considering case 2 and 3, we consider a classification line with a slope of 1.86 \citet[as in][]{nir}, but without shifting it by 0.05 mag. We sum all pixel values within the density map that satisfy this criterion to calculate the final fraction. Because each kernel is normalized to unit integral, this sum can be interpreted as the effective number of sources located within the NIR excess region after accounting for photometric uncertainties. We therefore can divide this number by the total number of sources to get the NIR excess fraction. 

This approach provides a probabilistic estimate of the NIR-excess population by explicitly incorporating the photometric uncertainties of individual sources. By propagating these uncertainties directly into the density map, the method avoids hard classifications of sources located near the NIR-excess boundary and instead estimates the expected number of objects within the selected region. This yields a more stable determination of the NIR-excess fraction in the presence of measurement uncertainties.

Following this method, we estimate the total fraction of NIR excess sources within our stellar sample:
\begin{enumerate}
    \item The large errors from the comparison with the 2MASS survey yield a global 27\% with NIR excess.
    \item The intermediate photometric errors reported in the HAWK-I manual produce a global 24\% with NIR excess.
\end{enumerate}

\begin{figure}[h]
    \centering
    \includegraphics[scale = 0.4]{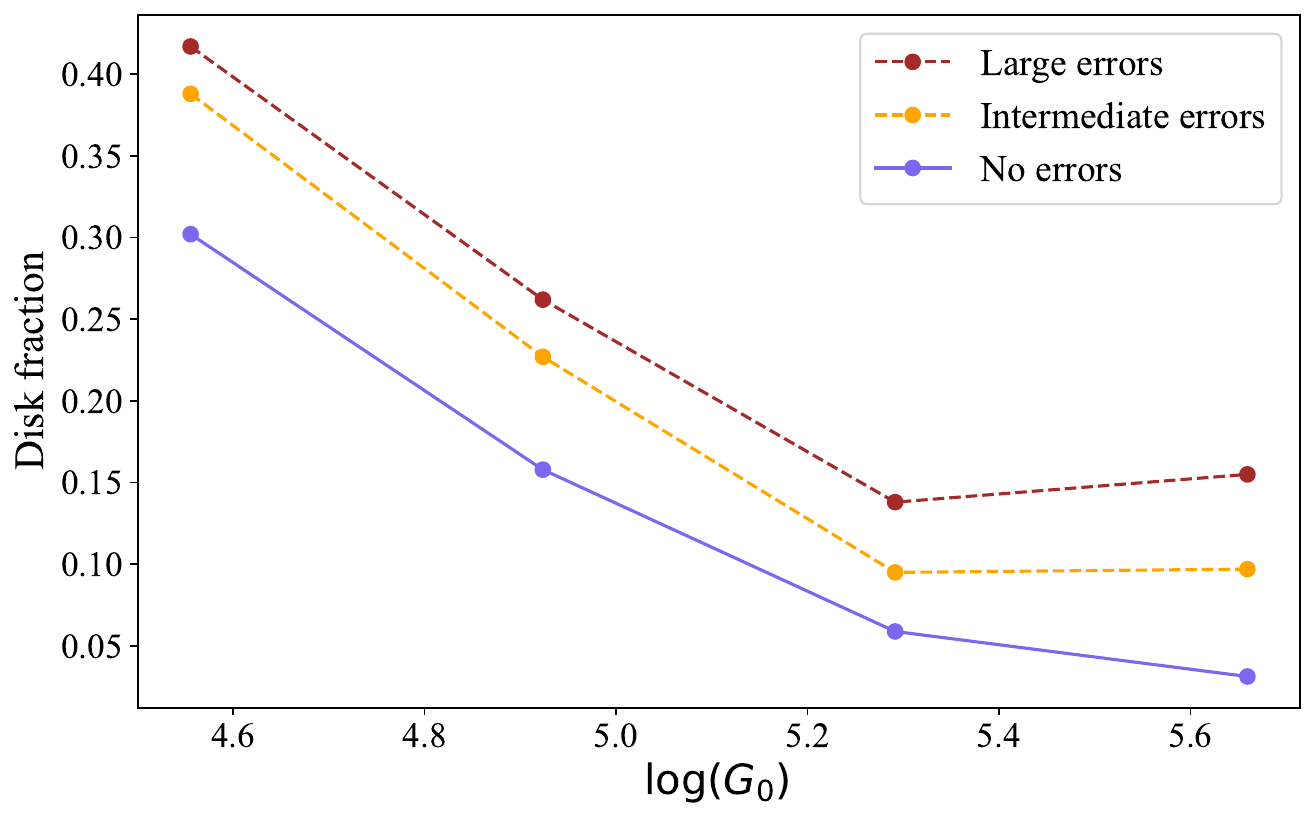}
    \caption{Disk fraction vs. FUV flux, estimated in 4 \go{} bins, comparing considering the errors or not. The FUV flux is displayed in the x-axis, in logarithmic scale and on units of \go{}. The fraction of stars with NIR excess estimated following the classification method by \citet{nir} is displayed as a purple curve. The fraction of stars with NIR excess estimated probabilistically, considering the intermediate errors in the HAWK-I manual, are displayed as an orange curve. Finally, the NIR excess fractions estimated using the same methodology, but considering the large errors from the comparison with 2MASS are displayed as a brown curve. \label{fig:disk_frac_new_errs}}
    
\end{figure}

Using this different classification method, we re-estimate the disk fractions in four FUV flux bins (same as in Fig.~\ref{fig:nir_frac}). We display the two distributions in Fig.~\ref{fig:disk_frac_new_errs} together with the disk fractions calculated in Sect.~\ref{sec:nir_go}, with the points positioned in the center of each bin.

We observe that, as expected, the trend of disk fraction vs. FUV flux follows the same behavior for all three curves: as the FUV flux increases, less sources exhibit a circumstellar disk. In addition, we derive systematically higher NIR excess fractions with the probabilistic approach, as it accounts for the non-uniform density distribution in the color-color diagram. Since the density of the points in the color-color diagram is not uniform, the classification which takes into account uncertainties will always provide larger fractions than the approach with no uncertainties. Overall, the fractions differ by up to approximately 15\%, depending on the adopted classification method, defining in that way a NIR excess range for each bin similar to the one when considering the symmetrical Poissonian errors. The only noticeable difference appears in the highest FUV bin, where the dashed curves show a slight increase in the NIR excess fraction. Specifically, the fractions increase by 2\% for the large errors and by 0.2\% for the intermediate errors. We attribute this behavior primarily to the limited statistics in this bin, which contains only 30 sources, half the size of the previous bin. Moreover, our classification method (purple curve) does not explicitly account for the individual photometric uncertainties of each source, which may lead to a slight underestimate of the number of NIR-excess sources. By contrast, the probabilistic approach incorporates larger photometric uncertainties together with the underlying density map, naturally resulting in somewhat higher estimated fractions.

\section{Sources contaminated by nearby O-type stars \label{app:weird_s}}

When constructing the HR diagram, we noted the presence of a group of 11 stars with $5500\text{ K} < T_{\text{eff}} < 7000 \text{ K}$ which do not follow a coherent sequence of evolutionary tracks predicted by the PARSEC models. We display this group in Fig.~\ref{fig:HR_weird} below. 

\begin{figure}[h!]
    \centering
    \includegraphics[scale = 0.4]{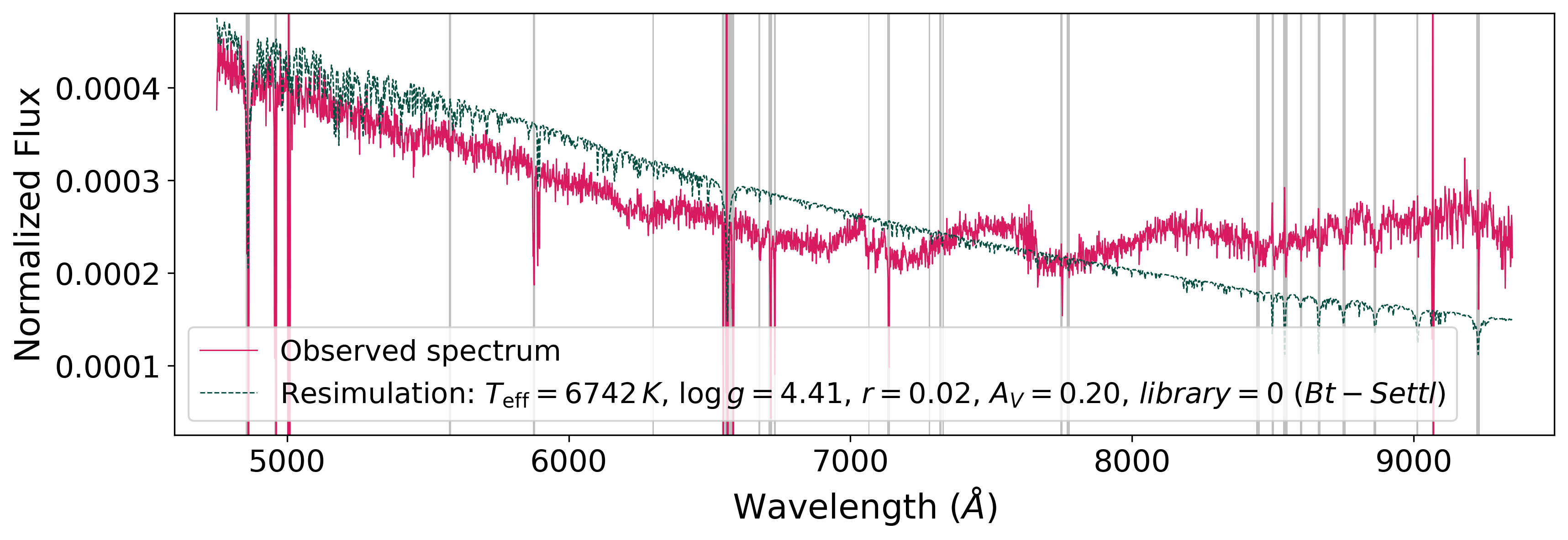}
    \caption{Reconstructed spectrum for one of the contaminated sources. The observed spectrum displayed in magenta for comparison.}
    \label{fig:resim_spec}
\end{figure}

When investigating the position of these sources in the MUSE images, we find that most of them are situated close to the O-type stars in the field, and their stellar spectra are heavily contaminated by the luminosity of these massive stars. We further investigate the rest by reconstructing their spectra using the SAPSAL estimated parameters and the synthetic spectral libraries of BT-Settl and Dusty. We find that the estimated parameters don't reconstruct the observed spectra (see fig.~\ref{fig:resim_spec} for an example). We therefore exclude these sources and restrict our final sample to 332 stars.

\begin{figure}[h!]
    \centering
    \includegraphics[scale = 0.38]{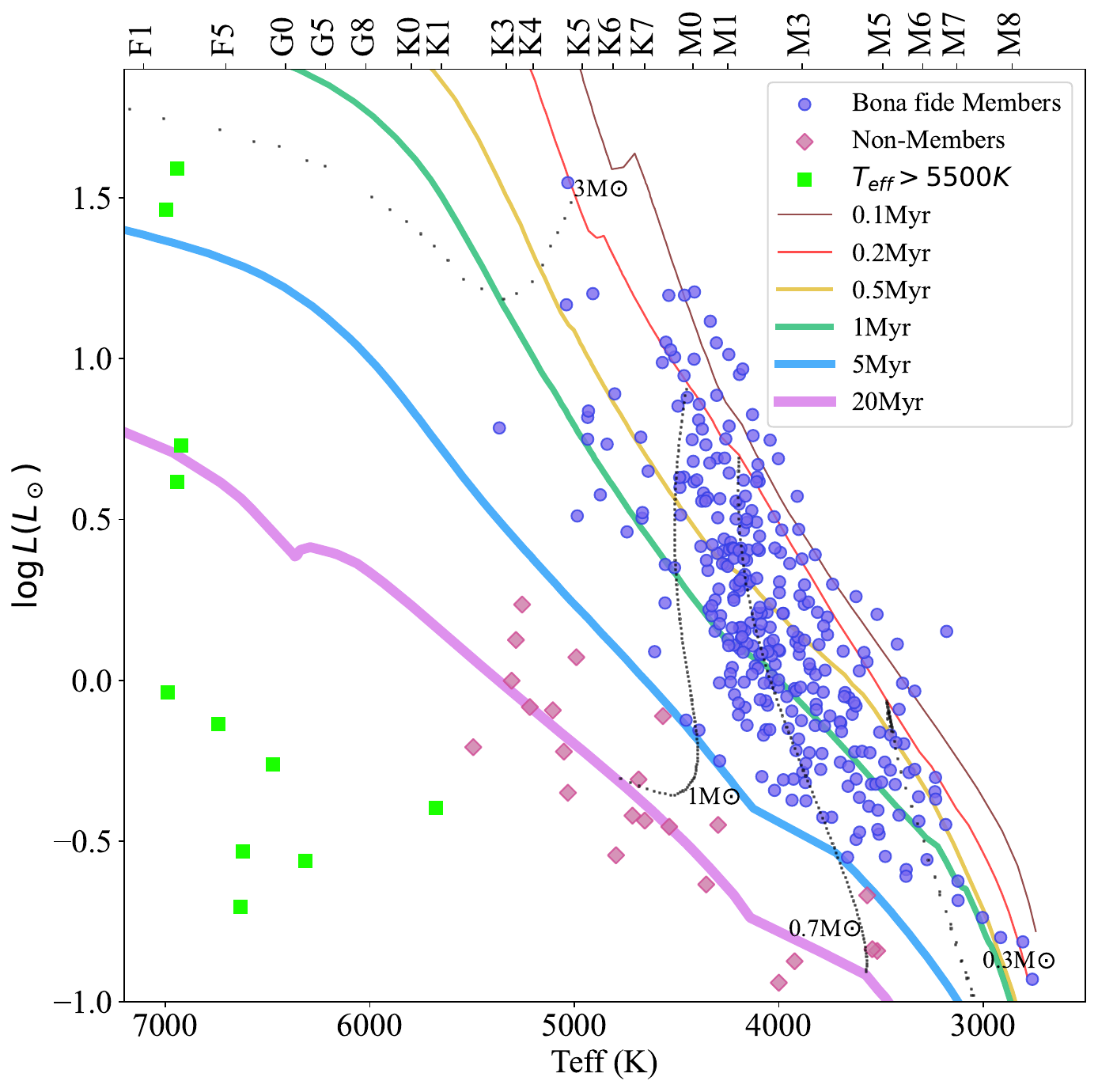}
    \caption{HR diagram with the initial sample from  Tr14's core. Blue circles represent the bona fide cluster members, whereas pink rhombuses represent stars considered as non-members. Green squares identify the non-coherent subpopulation excluded from the analysis. The evolutionary tracks from \citet{parsec} are shown as colorful lines indicating the isochrones; dotted gray lines indicate the isomasses.}
    \label{fig:HR_weird}
\end{figure}

\end{appendix}

\end{document}